\documentclass{elsart}
\usepackage[]{graphicx}
\usepackage{eufrak}
\usepackage{amsmath,epsf}
\newcommand{\bi}{{\boldsymbol{i}}}
\newcommand{\bj}{{\boldsymbol{j}}}
\newcommand{\bl}{{\boldsymbol{l}}}
\newcommand{\bp}{{\boldsymbol{p}}}
\begin{document}
\runauthor{Emig and Kardar}
\begin{frontmatter}
\title{Probability Distributions
of Line Lattices in Random Media from the 1D Bose Gas}
\author[1]{Thorsten Emig\thanksref{te}}\and
\author[1,2]{Mehran Kardar\thanksref{mk}}
\thanks[te]{E-mail: emig@mit.edu}
\thanks[mk]{E-mail: kardar@mit.edu}

\address[1]{Physics Department, Massachusetts Institute of Technology,\\
Cambridge, MA~02139, USA}

\address[2]{Institute for Theoretical Physics, University of California,\\ 
Santa Barbara, CA~93106, USA}

\begin{abstract}
The statistical properties of a two dimensional lattice of elastic
lines in a random medium are studied using the Bethe ansatz.  We
present a novel mapping of the dilute random line lattice onto the
weak coupling limit of a pure Bose gas with delta-function
interactions.  Using this mapping, we calculate the cumulants of the
free energy in the dilute limit exactly. The relation between density
and displacement correlation functions in the two models is examined
and compared with existing results from renormalization group and 
variational ans\"atze.

\end{abstract}
\end{frontmatter}

\section{Introduction}

Disorder, in the form of quenched impurities in a sample, is likely to
modify various measurements. However, from a theoretical point of
view, characterizing the behavior of disordered systems has proven
difficult, making this one of the most challenging and controversial
fields of statistical mechanics.  The directed polymer in a random
medium (DPRM) is one of the rare cases of a disorder-dominated system
whose statistical properties can be determined in great detail both
analytically and numerically \cite{Halpin-Healy95}, and thus serves as
a prototype for other such systems. Furthermore, its behavior is
related to a plethora of other problems in statistical mechanics, such
as the kinetic roughening of surfaces \cite{KPZ}, the hydrodynamics of
the Burger's equations \cite{Huse85}, and surprisingly the biological
sequence-alignment problem for genes and proteins \cite{Hwa96,Hwa99}.

The DPRM is directly relevant to vortex lines in superconductors with
point impurities. Most research on the vortex phases in these systems
has focused on the structural properties of the flux lattice in a
sample containing many vortices. Recently, the statistics of single
vortex lines was studied experimentally on a two dimensional flux line
lattice oriented parallel to a thin micrometer-sized film of
2H-NbSe$_2$ \cite{Bolle99}. Magnetic response measurements show
interesting sample dependent fine structure in the constitutive
relation $B(H)$, providing a fingerprint of the underlying random
pinning landscape. More generally, the random potential is expected to
modify the measurement of various thermodynamic and structural
quantities, making it desirable to characterize the probability
distribution functions (PDFs) for the outcome. At one extreme,
microscopic quantities such as two-point correlation functions are
quite sensitive to randomness, inducing in some cases complicated
multiscaling behavior of their PDFs \cite{Davis00}. On the other hand,
the free energy and other macroscopic quantities are expected to have
simpler self-averaging behavior, while non-trivial finite-size effects
are still present in their corresponding PDFs.

To date, the PDF of the free energy is known exactly only for a single
line, the DPRM \cite{Kardar87,Derrida88,Gorokhov99,Brunet00}. However,
as in superconductors, in most cases elastic lines appear at a finite
density, leading to novel effects resulting from a competition between
randomness and line interactions. Similar line lattice structures
occur in domain walls in the incommensurate phase of
charge-density-waves, in monolayers adsorbed on crystal surfaces, and
in the form of steps on the vicinal surface of a crystal
\cite{Mullins64}. As pointed out by de Gennes in the context of
polymers \cite{deGennes68}, and later on by Pokrovsky and Talapov for
domain walls in incommensurate phases \cite{Pokrovsky79}, the
statistical mechanics of a {\it pure} two-dimensional line lattice
maps onto one-dimensional free fermions if the repulsive line
interactions are replaced by non-crossing constraints. Applying the
same constraint, Kardar used the replica method to show that the line
lattice in a random potential maps onto one-dimensional $U(n)$
fermions with an attractive interaction \cite{Kardar87}. Calculating
the ground state energy of this quantum system for integer $n$ by
Bethe ansatz, and analytically continuing to $n \to 0$, the
quench-averaged free energy was calculated in this reference. Although
the PDF was not calculated explicitly, a scaling ansatz for the
replica free energy was proposed based on dimensional
arguments. Following up on this mapping, there have been attempts
\cite{Tsvelik92,Balents93} to calculate the density correlations of
the line lattice from the excitation spectrum of the $U(n)$ fermions
by Bethe ansatz, leading to different results.

In this paper we examine the analogy between the replicated line
lattice and $U(n)$ fermions further.  A considerable difficulty in the
replica approach is the need for analytical continuation to small $n$
of the results obtained for integer $n$. Analyticity is usually
assumed in replica theories, since it cannot be proved in most
cases. Here we develop a new method for analytical continuation in the
line lattice problem which is based on Euler's gamma function, the
natural continuation of the factorial to real valued arguments. This
technique provides an analytic expression for the replica free energy
of the line lattice at small $n$ in terms of an integral equation,
which in general has to be solved numerically. Interestingly, in the
dilute limit of the line lattice, the analytically continued Bethe
ansatz equations are identical to that of the 1D Bose gas with
delta-function repulsions. This novel mapping allows us to obtain
exact expressions for the free energy cumulants of the line lattice,
and to reveal new relations between density correlations in the 1D
Bose gas and the replica theory of the line lattice, respectively.
The scaling predictions for the replica free energy proposed in
\cite{Kardar87}, and used in \cite{Emig00} in a wider context of
disordered systems, are strikingly confirmed.

The outline of the paper is as follows.  Using the replica method and
the analogy between elastic lines and world-line of quantum particles,
we first show that the random line system is described by a quantum
problem of ``colored'' fermions, which interact upon contact. This
analogy is explained in detail in Section 2. Exact solutions for the
ground state energy of this colored fermion model in the thermodynamic
limit are then obtained via the series of nested Bethe ans\"atze
introduced by Yang \cite{Yang67}. For an integer number $n$ of colors,
the resulting sets of Bethe ansatz equations can be reduced to a
single set of equations by considering $n$-clusters of fermions, as
showed by Takahashi \cite{Takahashi70} and Kardar
\cite{Kardar87}. These integer $n$ equations are derived in Subsection
3.1. In the following Subsection 3.2, we develop a new technique to
continue the discrete Bethe ansatz equations directly to continuous
$n$. We find that for small $n$, and in the limit of low line density,
the complex $n$-color fermion problem can be mapped onto the
well-studied one-dimensional Bose gas with delta-function repulsions
\cite{Lieb63} (see Section 4). It turns out that the Bose gas
interaction strength is proportional to the number $n$ of replicas,
thus mapping the interesting regime $n\ll 1$ into the weak coupling
limit of the Bose gas. Therefore, the problem of calculating disorder
averaged cumulants of the free energy of the line system becomes
equivalent to a perturbative expansion of the ground state energy of
the one-dimensional Bose gas in the weak coupling limit. The exact
results for the cumulants of the free energy are presented in Section
5. Finally, in Section 6 we apply our novel mapping to the question of
what can be learned about the density-density correlation function of
the line lattice from the known results for corresponding correlation
functions of the 1D Bose gas.

\section{Mapping fluctuating lines to fermions}

Configurations of directed elastic lines can be regarded as the world
lines of quantum particles, where the coordinate along the line plays
the role of time. To develop a concrete picture of this analogy, we
consider the canonical partition sum $Z_N$ of a $1+1$ dimensional
lattice of $N$ interacting lines in a random environment of lateral
length $L$. To simplify the sum over all configurations of the lines,
we assume periodic boundary conditions in the direction parallel to
the lines.  This is justified since the statistical mechanics of the
line lattice is insensitive to the particular choice of boundary
conditions in the limit where $L$ is much larger then the spacing
between line collisions \cite{Nelson89}. Now the partition sum can be
written as the path integral
\begin{equation}
\label{orig_Z}
Z_N[V(x,y)]=\frac{{\mathcal Z}_0^N}{N!} \sum_P \prod_{j=1}^N
\int dr_j \int_{x_j(0)=r_j}^{x_j(L)=r_{P_j}} {\mathcal D}
x_j(y)\exp\left\{-H[V(x,y)]/T\right\},
\end{equation}
where we have summed over all permutations $P$ and have used Feynman's
definition of the path-integral measure ${\mathcal D} x_j(y)$
\cite{Feynman_path}. Here ${\mathcal Z}_0$ denotes the partition
function of a single line in a pure system with only one end of the
line fixed\footnote{${\mathcal Z}_0$ depends on the lattice constants
of the discretization of the path integral. The partition function of
the interacting random line system will be expressed relative to
${\mathcal Z}_0$ \cite{Nelson89}.}. The Hamiltonian in
Eq. (\ref{orig_Z}) is given by
\begin{equation}
H[V(x,y)]=\int_0^L dy \left\{ \sum_{j=1}^N \frac{g}{2}
\left(\frac{dx_j}{dy}\right)^2+ \sum_{i<j} U(x_i-x_j)+
\sum_{j=1}^N V(x_j,y) \right\},
\end{equation}
which is a functional of the random potential $V(x,y)$. The first term
measures the elastic energy of individual lines with line tension
$g$. The lines interact via a short ranged repulsive pair potential
$U(x)$, which is local in $y$. Effectively, a line interacts only with
the lines in the same constant $y$ cross section. This assumption is
valid if the line coordinates $x_j(y)$ vary slowly with $y$. The last
term couples the potential from randomly distributed impurities to the
local line positions, and makes the Hamiltonian ``time'' dependent.
Without this last term, the expression for $Z_N$ is formally identical
to the Feynman path integral for the canonical partition sum of an
interacting 1D Bose gas \cite{Nelson89}. But there is an additional
restriction for the configurations of elastic lines compared to the
boson world lines. Usually the overlap of lines is associated with a
high energy cost and, therefore, we treat the lines as non-crossing,
corresponding to an infinite potential $U(x)$ at $x=0$. As a
consequence, only the identity permutation contributes to the sum in
Eq. (\ref{orig_Z}).

The free energy $F[V(x,y)]=-T\ln Z_N$ is itself a random variable.  To
obtain the probability distribution function (PDF) of the free energy,
we use the replica method. Moments of the partition function are
obtained by introducing $n$ replicas of the original system, described
by line coordinates $x_{j,\alpha}(y)$, $\alpha=1,\ldots,n$, and then
averaging over the random potential. Assuming that $V(x,y)$ is
Gaussian distributed with zero mean and spatially uncorrelated with
variance
\begin{equation}
\label{randpot_corr}
[V(x,y)V(x',y')]=\Delta \delta(x-x')\delta(y-y'),
\end{equation}
where $[\ldots]$ denotes impurity averaging, the average over the
random potential yields
\begin{equation}
\label{rep_Z}
[Z_N^n]=\frac{{\mathcal Z}_0^{nN}}{(N!)^n}
\prod_{\alpha=1}^n \prod_{j=1}^N \int dr_{j,\alpha} 
\int_{x_{j,\alpha}(0)=r_{j,\alpha}}^{x_{j,\alpha}(L)=r_{j,\alpha}} 
{\mathcal D} x_{j,\alpha}(y)
\exp\left\{-H_n/T\right\},
\end{equation}
with the replicated Hamiltonian
\begin{eqnarray}
H_n&=&\int_0^L dy \left\{ \sum_{\alpha=1}^n \sum_{j=1}^N \frac{g}{2}
\left(\frac{dx_{j,\alpha}}{dy}\right)^2+ \sum_{\alpha=1}^n\sum_{i<j} 
U(x_{i,\alpha}(y)-x_{j,\alpha}(y))\right.\nonumber \\
&&-\left.\frac{\Delta}{2T}\sum_{\alpha,\beta=1}^n
\sum_{i,j=1}^N \delta(x_{i,\alpha}(y)-x_{j,\beta}(y))\right\}.
\end{eqnarray}
The sum over permutations has been eliminated from Eq. (\ref{rep_Z})
due to the non-crossing condition. The impurity averaged moments of
$Z_N$ are obtained from the time-independent Hamiltonian $H_n$ at the
cost of introducing a new attractive interaction proportional to the
variance $\Delta$ of the random potential. This attraction can be
understood as follows. In the original random system the lines prefer
to wander through regions of favorable values of the random potential
$V(x,y)$. Due to the impurity average the system becomes
translationally invariant and there are no longer energetically
favored regions of space-time. Instead, in calculating $n$th moments
of $Z_N$, the system of $n$ copies of the lines gains due to
$[\exp(-V(x,y))]= \exp(\Delta/2)$ an energy $\Delta/2T$ per crossing
of two lines (of different replicas), thus leading to an attraction.

The time-independence of the replicated Hamiltonian, and the choice of
periodic boundary conditions, allows us to write the partition function
in terms of the symmetric statistical density matrix
$\rho_{sym}(x_{1,1},\ldots,x_{N,n};x'_{1,1},\ldots,x'_{N,n};L)$ as an
integral over all boundary positions of the lines at $y=0$,
\begin{equation}
\label{Z_dm_rel}
[Z_N^n]={\mathcal Z}_0^{nN}\int \prod_{j=1}^N \prod_{\alpha=1}^n dx_{j,\alpha}
\rho_{sym}(x_{1,1},\ldots,x_{N,n};x_{1,1},\ldots,x_{N,n};L).
\end{equation}
The density matrix itself can be written as an imaginary time Feynman
path integral for the world lines of $Nn$ particles with fixed
boundary positions $x_{1,1},\ldots,x_{N,n}$ at $y=0$ and
$x'_{1,1},\ldots,x'_{N,n}$ at $y=L$ \cite{Feynman_path}. The density
matrix can alternatively be expressed in coordinate representation as
\begin{eqnarray}
\label{dm_rep}
&&\rho_{sym}(x_{1,1},\ldots,x_{N,n};x'_{1,1},\ldots,x'_{N,n};L)=
\nonumber\\
&&\quad=\sum_i^{sym} e^{-\beta_{\rm qm}E_i}\psi_i(x_{1,1},\ldots,x_{N,n})
\psi_i^*(x'_{1,1},\ldots,x'_{N,n}),
\end{eqnarray}
with the mapping $\beta_{\rm qm} \to \beta_{\rm cl}L=L/T$, $\hbar \to
\beta_{\rm cl}^{-1}=T$ between quantum mechanical and classical
parameters\footnote{Here and in the following, we set $k_B=1$.}. The
sum runs over all symmetric states with energy eigenvalues $E_i$ and
eigenfunctions $\psi_i$ of the 1-dimensional quantum-Hamiltonian
\begin{eqnarray}
\label{qm_ham}
&&\hat H= -\frac{T^2}{2g} \sum_{\alpha=1}^n \sum_{j=1}^N 
\frac{\partial^2}{\partial x_{j,\alpha}^2} + \sum_{\alpha=1}^n
\sum_{i<j} U(x_i-x_j) -\frac{\Delta}{2T} \sum_{\alpha,\beta=1}^n
\sum_{i,j=1}^N \delta(x_{i,\alpha}-x_{j,\beta}).\nonumber\\
\end{eqnarray}
From a classical statistical mechanics point of view, the operator
$\exp(-\hat H/T)$ is the transfer matrix corresponding to the
partition function $[Z_N^n]$.  

Using the representation of the density matrix in Eq. (\ref{dm_rep}),
we can perform the integrations in Eq. (\ref{Z_dm_rel}) easily. Taking
the thermodynamic limit $\beta_{\rm qm}\sim L \to \infty$, the free
energy $F_n$ of the replicated system is given by the ground state
energy $E_0$ of the quantum system in Eq. (\ref{qm_ham}),
\begin{equation}
\label{F_n}
F_n=-T\ln [Z_N^n] = -TnN\ln({\mathcal Z}_0) +E_0L.
\end{equation}
Therefore, the statistical properties of the line lattice in the
thermodynamic limit $L\to\infty$ are dominated by the ground state
wave function $\psi_0$, which must be of bosonic symmetry
\cite{Feynman_path}. To obtain the disorder averaged free energy $[F]$
of the original line system within the replica method usually the
identity $[\ln Z_N]=\lim_{n\to 0}([Z_N^n]-1)/n$ is used. However,
calculating the average $[Z_N^n]$ for all values $n>0$ by analytic
continuation of $E_0(n)$ yields much additional information about the
PDF of the free energy beyond $[F]$.  Since $[Z_N^n]$ is the
characteristic function of the random variable $[\ln Z]$, expansion
with respect to $n$ shows that
\begin{equation}
\label{Fcumulants}
nN\ln {\mathcal Z}_0 - \frac{E_0(n)L}{T} = \ln [Z_N^n] = \sum_{j=1}^\infty
\frac{(-n)^j}{j!}\frac{[F^j]_c}{T^j},
\end{equation}
where $[F^j]_c$ is the $j$th cumulant of the free energy. This
identification relies on the possibility of an analytic expansion
around $n \to 0$.  That this is indeed the case for the line system
considered here will be further discussed below.

Determining the PDF of the free energy is thus reduced to calculating
the ground state energy of the quantum-mechanical system in
Eq. (\ref{qm_ham}). Before we can proceed with the calculation of
$E_0(n)$, we have to specify the pair potential $U(x)$. We will
consider the limit where the characteristic length scale of the short
ranged pair potential goes to zero, leaving just a non-crossing
condition for the lines. This replacement of the interaction potential
by a pure geometric constraint for the line configurations does not
change the physical mechanisms involved in the non-trivial
characteristics of the line system.  The crucial competition between
the elasticity of single lines, the tendency of lines to follow energy
valleys of the random potential, and the restriction of configurations
by adjacent lines are preserved. Within the quantum-mechanical
description of the lines, there is a convenient way to implement the
non-crossing condition: As noted first by Pokrovsky and Talapov
\cite{Pokrovsky79} for pure systems, the Pauli exclusion principle can
be used to represent the elastic lines as world lines of fermions,
which automatically avoids crossings.  We thus look for a ground state
wave function $\psi_0$ with a suitable antisymmetry under particle
permutations, which describes $Nn$ particles with just an {\em
attractive} interaction on contact between {\em any} two particles. It
is interesting to note that we change the symmetry of the originally
bosonic ground state function by hand by applying the exclusion
principle. However, the ground state energy we are interested in is
the same both in the fermion and the impenetrable boson description.
Introducing the new subscript $\bi=(i-1)n+\alpha$ instead of ($i$,
$\alpha$) to number consecutively all $nN$ particles, the Hamiltonian
(\ref{qm_ham}) can now be written as
\begin{equation}
\label{bethe_ham}
\hat H =-\frac{\hbar^2}{2m}\sum_{\bi=1}^{nN}
\frac{\partial^2}{\partial x_\bi^2} -2c_0\sum_{\bi<\bj}
\delta(x_\bi-x_\bj),
\end{equation}
with mass $m=g$ and interaction amplitude $c_0=\Delta/2T$.  The
explicit distinction between lines in the same or different replicas
has disappeared from the Hamiltonian, but is hidden in the required
transformation properties of the wave function under particle
permutations.  The appropriate symmetry of the ground state wave
function is restricted by two conditions: (i) $\psi_0$ must be {\em
antisymmetric} if two of the $N$ particles in the {\em same} replica
(color) are interchanged (Fermi statistics). (ii) $\psi_0$ must be
{\em symmetric} if two of the $n$ {\em different} replica (colored)
particles corresponding to the same line are exchanged since replicas
are equivalent. Due to its antisymmetry, $\psi_0$ tends to zero
if two particles of the same color approach each other and the
attractive delta-function potential becomes inactive. This guarantees
automatically that the impurity averaging results effectively in an
attractive interaction on contact between particles of {\em different}
color only. Therefore, the sum over pairs of lines interacting via a
delta-function in Eq. (\ref{qm_ham}) yields no contribution if
$\alpha=\beta$, resulting in the sum over all pairs in
Eq. (\ref{bethe_ham}). To conclude, the original quantum Hamiltonian
(\ref{qm_ham}) with $U(x)$ replaced by the non-crossing condition for
lines belonging to the same replica is equivalent to the Hamiltonian
(\ref{bethe_ham}), if one solves for a ground state wave function of
appropriate symmetry.

\section{The Bethe Ansatz}

In one dimension, a number of interacting quantum systems can be
solved exactly by the Bethe ansatz \cite{Bethe31,Thacker81}. Lieb and
Liniger were the first to apply the Bethe ansatz method to a continuum
quantum field theory by solving for the ground state energy and the
excitation spectrum of the one dimensional interacting Bose gas with
repulsive delta-function pair potentials \cite{Lieb63}. While their
Hamiltonian is exactly given by Eq. (\ref{bethe_ham}), the ground
state wave function that we are looking for has a more complicated
structure since it has to describe fermions of $n$ different
colors. The case $n=2$ corresponds to spin-$\frac{1}{2}$ fermions and
was solved some time ago by Gaudin \cite{Gaudin67} for attractive
interactions, and by Yang \cite{Yang67} for repulsive
interactions. Choosing $n=3$ colors, we obtain a model for nuclear
matter with an attractive quark-quark interaction, which has been
discussed in connection with quantum chromodynamics and also studied
by Bethe ansatz methods \cite{Koltun87}. Since we are interested in an
analytical continuation of the ground state energy for small $n\ll 1$,
first we find a solution for integer $n$ and then continue it
analytically. According to the symmetry requirements discussed above,
the allowed spatial wave functions of the $U(n)$ fermions have to
transform like the irreducible representation of the symmetric group
$\mathfrak S_{nN}$ corresponding to the Young diagram $[n^N]$, which
consists of $N$ rows of equal length $n$. Sutherland has constructed a
general solution of the eigenvalue problem for Hamiltonian
(\ref{bethe_ham}) with $c_0<0$ when the wave function transforms like
an arbitrary irreducible representation of the symmetric group
\cite{Sutherland68}. The method involves a nested series of Bethe
ans\"atze and leads to coupled integral equations for momentum
densities. The same method has been applied later to the attractive
case by Takahashi \cite{Takahashi70} and Kardar \cite{Kardar87},
leading to considerable simplifications in the solution of the Bethe
ansatz equations. Before we turn to the interesting regime $n\ll 1$,
we explain briefly the idea of the method of nested Bethe ans\"atze,
in a simplified version for the relevant irreducible representation
$[n^N]$ with integer values $n\ge 1$.

\subsection{Bethe Ansatz equations for integer $n$}
\label{sec:bethe}

The eigenstates of the Hamiltonian (\ref{bethe_ham}) are given by
Bethe's hypothesis \cite{Bethe31}. For $0<x_{Q_1}<\ldots <
x_{Q_{nN}}<W$, with $W$ the transversal system size, we can write the
wave function as
\begin{equation}
\label{wavefunction}
\Psi=\sum_{\mathcal P} [Q,P] \exp\left(i\sum_{\bj=1}^{nN} \lambda_{P_\bj}
x_{Q_\bj} \right). 
\end{equation}
Here $Q$ and $P$ are permutations in $\mathfrak S_{nN}$, and the
numbers $[Q,P]$ form a $(nN)!\times (nN)!$ matrix.  In the following
we denote the columns of this matrix by the vector $\xi_P=[\,\cdot
,P]$. Next, we define the permutation operators ${\mathcal P}_{lm}$,
which act on the $\xi_P$. Their effect is to interchange the $Q_l$
component with the $Q_m$ component of the vector $\xi_P$, i.e.,
${\mathcal P}_{lm} [Q,P] \equiv [Q',P]$ where the new permutation $Q'$
is defined by $Q'_l=Q_m$, $Q'_m=Q_l$, and $Q'_j=Q_j$ for $j\neq
l,m$. These operators form a $(nN)!$ dimensional representation of the
symmetric group ${\mathfrak S}_{nN}$. But this representation is
reducible, i.e., for a wave function of a given symmetry, many
coefficients $[Q,P]$ are identical.  Therefore, in the following we
consider implicitly only the subspace of the original $\xi_P$-space
which corresponds to the irreducible representation $[n^N]$ of
${\mathfrak S}_{nN}$. On this subspace the operators ${\mathcal
P}_{lm}$ act as matrices of the dimension corresponding to the
irreducible representation.

The requirement of a continuous wave function, which has cusps
wherever two of its coordinates are equal in order to compensate the
delta-function interaction in Eq. (\ref{bethe_ham}), imposes relations
between the coefficients $[Q,P]$. In terms of two vectors $\xi_P$,
$\xi_{P'}$ corresponding to permutations $P$ and $P'$, respectively,
which differ by a transposition of $i$ and $j$, i.e, $P_l=i=P'_m$,
$P_m=j=P'_l$, and $P_j=P'_j$ for $j\neq l,m$, the relations can be
written as
\begin{equation}
\label{xi-rel-1}
\xi_{P'}=Y_{ij}^{lm} \xi_P,
\end{equation}
with the operator
\begin{equation}
Y_{ij}^{lm}=\frac{ic}{\lambda_i-\lambda_j-ic} 
+\frac{\lambda_i-\lambda_j}{\lambda_i-\lambda_j-ic}{\mathcal P}_{lm},
\end{equation}
and the rescaled interaction strength
$c=2mc_0/\hbar^2=g\Delta/T^3$. If these relations are fulfilled, the
wave function (\ref{wavefunction}) is an eigenstate of the Hamiltonian
(\ref{bethe_ham}) for {\em any} set of momenta $\lambda_i$. The
allowed momenta are then restricted by imposing periodic boundary
conditions on the wave function. Expressed in terms of $\xi_I$
corresponding to the identity element $I$ of ${\mathfrak S}_{nN}$, the
conditions force $\xi_I$ to be simultaneously an eigenvector of the
$nN$ matrix equations
\begin{equation}
e^{i\lambda_j W}\xi_I=X_{j+1\,j}X_{j+2\,j}\cdots X_{N\,j}X_{1\,j}\cdots
X_{j-1\,j}\xi_I,
\end{equation}
with $j=1,\ldots,nN$ and $X_{ij}(\{\lambda_j\}) \equiv{\mathcal
P}_{ij} Y^{ij}_{ij}$.  The $\lambda_j$ which solve these equations
depend, of course, on the chosen irreducible representation.

So far, we have considered only the spatial part (\ref{wavefunction})
of the full fermionic wave function. The full wave function is given
by the product of (\ref{wavefunction}) and the color wave function
$\chi$.  The product has to be antisymmetric under simultaneous
interchange of both position and color of two particles. To guaranty
this property, the color wave function $\chi$ has to transform like
the conjugate irreducible representation $[N^n]$. One can think of
the basis vectors of this representation as all the allowed color
sequences of length $nN$ which form by linear combination the function
$\chi$. Now, the crucial point of Yang's and Sutherland's method for
determining the momenta $\lambda_i$ is to solve the matrix equations
in color space by regarding the color wave function as a spatial wave
function describing $(n-1)N$ distinguishable particles on a {\em
discrete} cyclic chain with $N$ identical vacancies. By making a
generalized Bethe ansatz introduced by Yang, this new problem can be
cast into matrix equations identical in form to the original one,
but corresponding to the lower dimensional representation $[N^{n-1}]$
for the color wave function $\chi$.  This procedure is then continued
until the problem has been reduced to one for particles of a single
color only. To be more concrete, the periodic boundary conditions for
$\chi$ read
\begin{equation}
\label{pbc_chi}
\mu_j \chi=X'_{j+1\,j}X'_{j+2\,j}\cdots
X'_{N\,j}X'_{1\,j}\cdots X'_{j-1\,j}\chi,
\end{equation}
where
\begin{equation}
\label{ev_mu}
\mu_j=e^{i\lambda_j W},
\end{equation}
and $X'_{ij}(\{\lambda_j\})$ is obtained from $X_{ij}(\{\lambda_j\})$
by the replacement ${\mathcal P}_{ij} \to -{\mathcal P}_{ij}$.  This
change of sign for the permutation operator comes from the fact that
in general two representation matrices of a permutation $P$
corresponding to the original and conjugated representation,
respectively, differ by the parity of $P$ only.  To solve
Eqs. (\ref{pbc_chi}) we use the generalized Bethe hypothesis of Yang
for $\chi$ on the discrete cyclic chain. It consists in the ansatz
\begin{equation}
\label{Yang-hypo}
\chi=\sum_P [Q,P] G(\lambda^{(2)}_{P_1},y_{Q_1})\cdots
G(\lambda^{(2)}_{P_{(n-1)N}},y_{Q_{(n-1)N}}),
\end{equation}
where the integers $1\le y_{Q_1} < \ldots < y_{Q_{(n-1)N}} \le nN$
denote the coordinates of the distinguishable particles, and
$\lambda^{(2)}_1,\dots,\lambda^{(2)}_{(n-1)N}$ is a set of unequal
complex quasi-momenta. $P$ and $Q$ are now elements of $\mathfrak
S_{(n-1)N}$. The coefficients $[Q,P]$ form now a $(n-1)N \times
(n-1)N$ matrix, the columns of which we denote by $\xi'_P$. Again, we
define permutation operators $\hat {\mathcal P}_{lm}$ that act on
$\xi'_P$ so that they interchange $Q_l$ and $Q_m$. By considering a
suitable subspace of the $\xi'_P$-space, these operators are chosen to
form the irreducible representation $[N^{n-1}]$ of $\mathfrak
S_{(n-1)N}$, in order to assure that the color wave function has the
appropriate symmetry. Physically, the number $\chi$ associated with a
particular set $\{y_1,\ldots,y_{(n-1)N}\}$ and permutation $Q$ can be
interpreted as the amplitude for finding a particular arrangement
of colors for the $nN$ particles. Indeed, the positions of all the
particles of one color, say red, determine the $y_i$ of the remaining
particles, and the color arrangement of these remaining $(n-1)N$
particles can be identified with the permutation $Q$. To complete the
ansatz, we have to specify the function $G(\lambda,y)$,
\begin{equation}
G(\lambda,y)=\prod_{j=1}^{y-1} \frac{i(\lambda_j-\lambda)+c/2}
{i(\lambda_{j+1}-\lambda)-c/2}.
\end{equation}
The requirement that this ansatz for $\chi$ is a simultaneous
eigenvector of the $nN$ matrix equations (\ref{pbc_chi}) imposes
relations between the $\xi'_P$, as the original eigenvalue problem for
the Hamiltonian (\ref{bethe_ham}) does for the $\xi_P$. These
relations can again be written as
\begin{equation}
\label{xi-rel-2}
\xi'_{P'}={Y'}_{lm}^{ij}\xi'_P,
\end{equation}
with the operator 
\begin{equation}
{Y'}_{lm}^{ij}=\frac{-ic}{\lambda^{(2)}_i-\lambda^{(2)}_j+ic} 
+\frac{\lambda^{(2)}_i-\lambda^{(2)}_j}{\lambda^{(2)}_i-\lambda^{(2)}_j+ic}
\hat {\mathcal P}_{lm},
\end{equation}
and $P$, $P'$ defined as above Eq. (\ref{xi-rel-1}). Finally, we have
to impose periodic boundary conditions on $\chi$. Using the relations
(\ref{xi-rel-2}), we obtain for the vector $\xi'_I$ the $(n-1)N$
matrix equations
\begin{equation}
\label{maeqs1}
\prod_{j=1}^{nN} \frac{i(\lambda^{(2)}_l-\lambda_j)-c/2}
{i(\lambda^{(2)}_l-\lambda_j)+c/2}\xi'_I=
X''_{l+1\, l}\cdots X''_{(n-1)N\, l}X''_{1\,l}\cdots X''_{l-1\,l}\xi'_I,
\end{equation}
for $l=1,\ldots,(n-1)N$, and with 
\begin{equation}
\label{defX2}
X''_{ij}(\{\lambda^{(2)}_l\})\equiv \hat{\mathcal P}_{ij}{Y'}_{ij}^{ij}=
\frac{\lambda^{(2)}_i-\lambda^{(2)}_j-ic}{\lambda^{(2)}_i-\lambda^{(2)}_j+ic}
X'_{ij}(\{\lambda^{(2)}_l\}).
\end{equation}
Substituting Eq. (\ref{defX2}) in Eq. (\ref{maeqs1}), we obtain new
matrix equations, which parallel in form precisely the original
equations (\ref{pbc_chi}) for the color wave function, but with
operators $X'_{ij}$, in which the quasi-momenta $\lambda^{(2)}_l$ have
replaced the original momenta $\lambda_l$. The new equations read
\begin{equation}
\label{pbc_chi_2}
\mu'_l \xi'_I = X'_{l+1\, l}\cdots X'_{(n-1)N\, l}X'_{1\,l}
\cdots X'_{l-1\,l}\xi'_I,
\end{equation}
with the new set of $\mu'_l$ defined via
\begin{equation}
\label{mu_p}
-\mu'_l \prod_{m=1}^{(n-1)N}
\frac{i(\lambda^{(2)}_l-\lambda^{(2)}_m)-c}
{i(\lambda^{(2)}_l-\lambda^{(2)}_m)+c}=\prod_{j=1}^{nN}
\frac{i(\lambda^{(2)}_l-\lambda_j)-c/2}{i(\lambda^{(2)}_l-\lambda_j)+c/2}.
\end{equation}
Thus Eq. (\ref{pbc_chi}) is reduced to the lower dimensional
Eq. (\ref{pbc_chi_2}), and the eigenvalue $\mu_j$ in Eq. (\ref{ev_mu}),
which is obtained from the ansatz (\ref{Yang-hypo}), is 
\begin{equation}
\label{mu_j}
\mu_j=\prod_{m=1}^{(n-1)N}\frac{i(\lambda_j-\lambda^{(2)}_m)+c/2}
{i(\lambda_j-\lambda^{(2)}_m)-c/2}.
\end{equation}
To reduce Eq. (\ref{pbc_chi_2}) further and to determine the
eigenvalue $\mu'_l$, we apply the procedure described above again in
the representation $[N^{n-2}]$, introducing the quasi-momenta
$\lambda^{(3)}_j$. This process terminates after $n-1$ iterations and
yields Sutherland's Bethe ansatz equations for the $Nn(n+1)/2$ complex
momenta $\lambda^{(\alpha)}_\bi$, with $\alpha=1,\ldots,n$ and
$\bi=1,\ldots,(n+1-\alpha)N$. Here we have denoted the original
momenta $\lambda_j$ by $\lambda^{(1)}_j$. 

Comparing Eq. (\ref{ev_mu}) with Eq. (\ref{mu_j}), we obtain
\begin{equation}
\label{nested_bethe_eqs_1}
e^{i\lambda^{(1)}_\bi W}=\prod_{\bj=1}^{(n-1)N}
\frac{i(\lambda^{(1)}_\bi-\lambda^{(2)}_\bj)+c/2}
{i(\lambda^{(1)}_\bi-\lambda^{(2)}_\bj)-c/2},\quad \text{with} 
\quad (\bi=1,\ldots,nN).
\end{equation}
Continuing the iteration, we obtain from Eq. (\ref{mu_p}) with
$\mu'_l$ given by Eq. (\ref{mu_j}), with $\lambda_j$ and
$\lambda^{(2)}_m$ replaced by corresponding higher order
quasi-momenta, the set of equations
\begin{equation}\begin{split}
\label{nested_bethe_eqs_2}
&\prod_{\bj=1}^{mN}\frac{i(\lambda^{(n-m+2)}_\bi-\lambda^{(n-m+1)}_\bj)-c/2}
{i(\lambda^{(n-m+2)}_\bi-\lambda^{(n-m+1)}_\bj)+c/2}=
-\prod_{\bl=1}^{(m-1)N}
\frac{i(\lambda^{(n-m+2)}_\bi-\lambda^{(n-m+2)}_\bl)-c}
{i(\lambda^{(n-m+2)}_\bi-\lambda^{(n-m+2)}_\bl)+c}\\
\quad & \times \prod_{\bp=1}^{(m-2)N}
\frac{i(\lambda^{(n-m+2)}_\bi-\lambda^{(n-m+3)}_\bp)+c/2}
{i(\lambda^{(n-m+2)}_\bi-\lambda^{(n-m+3)}_\bp)-c/2}, \quad
\text{with} \quad \begin{pmatrix} \bi=1,\ldots,(m-1)N\\ m=n,\ldots,3
\end{pmatrix}.
\end{split}\end{equation}
Finally, the iteration ends at the representation $[N]$, which
corresponds to $X'_{ij}=1$ in Eq. (\ref{pbc_chi_2}). Therefore, this
equation is trivially fulfilled with $\mu'_l=1$, corresponding
together with Eq. (\ref{mu_p}) to
\begin{equation}
\label{nested_bethe_eqs_3}
\prod_{\bj=1}^{2N} \frac{i(\lambda^{(n)}_\bi-\lambda^{(n-1)}_\bj)-c/2}
{i(\lambda^{(n)}_\bi-\lambda^{(n-1)}_\bj)+c/2}=-\prod_{\bl=1}^N
\frac{i(\lambda^{(n)}_\bi-\lambda^{(n)}_\bl)-c}
{i(\lambda^{(n)}_\bi-\lambda^{(n)}_\bl)+c}, \quad\text{with}\quad
(\bi=1,\ldots,N).
\end{equation}
In the following, we will make a simple ansatz
\cite{Takahashi70,Kardar87} for the complex momenta
$\lambda^{(\alpha)}_\bi$, which solve
Eqs. (\ref{nested_bethe_eqs_1})-(\ref{nested_bethe_eqs_3}) in the limit
$W\to\infty$, leaving only one set of $N$ independent equations for the
$N$ different real parts of the original momenta
$\lambda_\bi\equiv\lambda^{(1)}_\bi$. To motivate this ansatz, we
consider for the time being the case of a single line ($N=1$) coming
in $n$ colors. This corresponds to the representation $[n]$ for $n$
bosons without spin. Under the attractive interaction they form a
bound cluster, which is described by complex momenta
$\lambda_\alpha=ic(n+1-2\alpha)/2$ ($\alpha=1,\ldots,n$) forming a
so-called ``$n$ string'' \cite{Thacker81}. Switching back to the
general case of $N$ lines, but now for vanishing disorder strength
$\Delta$ so that $c=0$, the problem becomes that of $N$ free fermions
without spin ($n=1$) represented by the Young diagram $[1^N]$. In this
case, the momenta are equally spaced along the real axis between
$-k_F$ and $k_F=\pi \rho$, where $\rho=N/W$ is the density of
particles. With this two limiting cases in mind, it is reasonable to
make the general ansatz for large $W$
\begin{eqnarray}
\label{lam_ansatz_1}
\lambda^{(n)}_j&=&k_j +i B_{j,1}^{(n)}\quad (j=1,\ldots,N),\\
\label{lam_ansatz_2}
\lambda^{(n-m)}_{\bi=(j-1)(m+1)+\alpha}&=&k_j+ic\left(1-\alpha
+\frac{m}{2}\right) + i B_{j,\alpha}^{(n-m)} \\
&&\quad   (j=1,\ldots,N; \alpha=1,\ldots, m+1; m=1,\ldots,n-2),\nonumber\\
\label{lam_ansatz_3}
\lambda^{(1)}_{\bi=(j-1)n+\alpha}&=&k_j+ic\left(1-\alpha
+\frac{n-1}{2}\right) +i B_{j,\alpha}^{(1)}\\
&&\quad   (j=1,\ldots,N; \alpha=1,\ldots, n),\nonumber
\end{eqnarray}
where the $k_j$ are {\em real} valued momenta and the correction terms
$B_{j,\alpha}^{(n-m)}$ will be proven to vanish in the limit
$W\to\infty$ below. In constructing this ansatz we start with the
quasi-momenta $\lambda^{(n)}_i$, which were introduced in the last
iteration step of the nested Bethe ansatz. In this last step the
original problem of $nN$ interacting particles was completely reduced
to that of $N$ colorless cluster fermions. Therefore, in analogy to
the usual fermionic case discussed above, we make the ansatz that the
quasi-momenta $\lambda^{(n)}_i$ correspond to the real valued cluster
momenta $k_j$. But due to the complex structure of the effective
cluster-cluster interaction, the momenta $k_j$ are, of course, no
longer homogeneously distributed. To get a physical picture of the
internal structure of a single cluster of fermions, we think of it as
an ``$n$ string'' composed of $n$ bosons which are hold together by
the attractive interaction. From the above discussion of the $N=1$
case we know that the ``internal'' boson momenta of the cluster are
equally spaced along the imaginary axis. Therefore, the single cluster
wave function is localized in space within a range of size
$1/c=T^3/g\Delta$. The ans\"atze
(\ref{lam_ansatz_1})-(\ref{lam_ansatz_3}) show that also for a finite
cluster density in the limit $W\to\infty$, the overlap of clusters can
be neglected, particularly with regard to the internal structure of
the clusters.  Thus each cluster's momenta are simply given by the
momenta of an unperturbed ``$n$ string'' plus the collective real
valued cluster momentum $k_j$. This leads to the $nN$ original momenta
$\lambda^{(1)}_\bi$ of Eq. (\ref{lam_ansatz_3}). The intermediate
quasi-momenta of Eq. (\ref{lam_ansatz_2}) can be understood as that of
smaller sub-clusters which appear during the iterative scheme of the
nested Bethe ans\"atze.

We now prove that the ans\"atze
(\ref{lam_ansatz_1})-(\ref{lam_ansatz_3}) indeed solve
Eqs. (\ref{nested_bethe_eqs_1})-(\ref{nested_bethe_eqs_3}) in the
limit $W\to\infty$, leaving a new set of only $N$ equations to
determine the $k_j$'s. In substituting the ansatz in
Eqs. (\ref{nested_bethe_eqs_1})-(\ref{nested_bethe_eqs_3}) we keep the
correction terms $B_{j,\alpha}^{(n-m)}$ only in factors which
would become zero otherwise. The resulting equations for the
correction terms have two important properties: (i) From the first set
of $nN$ equations [Eq. (\ref{nested_bethe_eqs_1})] one observes that
the $B_{j,\alpha}^{(n-m)}$ have to vanish exponentially fast if
$W\to\infty$. (ii) The solutions depend formally on the real momenta
$k_j$, which are still free parameters. In the following we will show
that these $k_j$ have to fulfill necessarily $N$ independent equations
in order to make the equations for the $B_{j,\alpha}^{(n-m)}$
consistent and thus solvable.

Multiplying the $n$ equations of Eq. (\ref{nested_bethe_eqs_1}) with
the ansatz substituted and $\bi=(i-1)n+\alpha$ with $i$ fixed and
$\alpha=1,\ldots, n$, we obtain
\begin{equation}
\label{set1_reduced}
e^{ik_i nW}=(-1)^{n-1}\prod_{\beta=1}^{n-1}
\frac{B_{i,\beta}^{(1)} - B_{i,\beta}^{(2)}}
{B_{i,\beta+1}^{(1)} - B_{i,\beta}^{(2)}}\prod_{\substack{j=1\\j\neq i}}^N
\prod_{\alpha=1}^{n-1}\frac{i(k_i-k_j)+\alpha c}{i(k_i-k_j)-\alpha c}.
\end{equation}
In the next step we eliminate all correction terms
$B_{i,\alpha}^{(1)}$, $B_{i,\alpha}^{(2)}$ from this equation without
calculating them explicitly.  Thus we substitute the ansatz into
the set of Eqs. (\ref{nested_bethe_eqs_2}). The cluster momenta $k_j$
drop out and we get for $m=3,\ldots, n$; $i=1,\ldots, N$, and
$\beta=2,\ldots,m-2$, the new set of equations
\begin{equation}
\label{Bcoeff_rel_1}
-\frac{B_{i,\beta}^{(n-m+2)}-B_{i,\beta}^{(n-m+1)}}
{B_{i,\beta}^{(n-m+2)}-B_{i,\beta+1}^{(n-m+1)}}
=
\frac{B_{i,\beta}^{(n-m+2)}-B_{i,\beta-1}^{(n-m+2)}}
{B_{i,\beta}^{(n-m+2)}-B_{i,\beta+1}^{(n-m+2)}}\cdot
\frac{B_{i,\beta}^{(n-m+2)}-B_{i,\beta}^{(n-m+3)}}
{B_{i,\beta}^{(n-m+2)}-B_{i,\beta-1}^{(n-m+3)}}.
\end{equation}
In the special case $\beta=1$ we have
\begin{equation}
\label{Bcoeff_rel_2}
-\frac{B_{i,1}^{(n-m+2)}-B_{i,1}^{(n-m+1)}}
{B_{i,1}^{(n-m+2)}-B_{i,2}^{(n-m+1)}}
=
\frac{B_{i,1}^{(n-m+2)}-B_{i,1}^{(n-m+3)}}
{B_{i,1}^{(n-m+2)}-B_{i,2}^{(n-m+2)}},
\end{equation}
and similarly for $\beta=m-1$,
\begin{equation}
\label{Bcoeff_rel_3}
-\frac{B_{i,m-1}^{(n-m+2)}-B_{i,m-1}^{(n-m+1)}}
{B_{i,m-1}^{(n-m+2)}-B_{i,m}^{(n-m+1)}}
=
\frac{B_{i,m-1}^{(n-m+2)}-B_{i,m-2}^{(n-m+2)}}
{B_{i,m-1}^{(n-m+2)}-B_{i,m-2}^{(n-m+3)}}.
\end{equation}
Now we take for fixed $m\in\{3,\ldots,n\}$ and $i\in\{1,\ldots,N\}$ the
product of the $m-1$
Eqs. (\ref{Bcoeff_rel_1})-(\ref{Bcoeff_rel_3}). Introducing the
following ratio of correction terms,
\begin{equation}
\label{calBdef}
{\mathcal B}_i(m)=\prod_{\beta=1}^{m-1}
\frac{B_{i,\beta}^{(n-m+2)}-B_{i,\beta}^{(n-m+1)}}
{B_{i,\beta}^{(n-m+2)}-B_{i,\beta+1}^{(n-m+1)}},
\end{equation} 
we obtain the simple recursion relation
\begin{equation}
{\mathcal B}_i(m)=-{\mathcal B}_i(m-1).
\end{equation} 
From the definition of ${\mathcal B}_i(m)$ it is easy to realize that
the coefficient of the right hand side of Eq. (\ref{set1_reduced}) is
given by $(-1)^{n-1}{\mathcal B}_i(n)=-{\mathcal B}_i(2)$. Finally, to
obtain ${\mathcal B}_i(2)$ we substitute the ans\"atze of
Eqs. (\ref{lam_ansatz_1})-(\ref{lam_ansatz_3}) into
Eq. (\ref{nested_bethe_eqs_3}) keeping again the correction terms only
in factors which would become zero otherwise. For $i=1,\ldots,N$,
we thus obtain
\begin{equation}
{\mathcal B}_i(2)=\frac{B_{i,1}^{(n)}-B_{i,1}^{(n-1)}}
{B_{i,1}^{(n)}-B_{i,2}^{(n-1)}}=-1.
\end{equation}
This result shows that the correction terms dependent coefficient of
Eq. (\ref{set1_reduced}) has to be one in order to have a consistent
set of equations for the $B_{j,\alpha}^{(n-m)}$. This in turn
determines the allowed real parts $k_j$ of the original momenta
$\lambda_\bi\equiv\lambda^{(1)}_\bi$.  They have to fulfill
Eq. (\ref{set1_reduced}), which now reads
\begin{equation}
\label{int_bethe_eq}
e^{ink_jW}=\prod_{\substack{l=1\\l\neq j}}^N \prod_{\alpha=1}^{n-1}
\frac{i(k_j-k_l)/c+\alpha}{i(k_j-k_l)/c-\alpha} \quad (j=1,\ldots,N).
\end{equation}
These are the final Bethe Ansatz equations\footnote{ The same
equations have been obtained for the case $n=3$ in the context of
quantum chromodynamics \cite{Koltun87} by a more direct approach
which does not use Sutherland's nested series of Bethe
ans\"atze. Instead it was a priori assumed that the overlap of
clusters can be neglected in calculating the momenta in the
thermodynamic limit.  }  which have to be solved to obtain the ground
state energy of the quantum Hamiltonian $\hat H$ in
Eq. (\ref{bethe_ham}) in the thermodynamic limit, as
\begin{equation}
\label{gsenergy}
E_0=\frac{\hbar^2}{2m}\sum_{\bi=1}^{nN} \lambda_\bi^2 =
\frac{\hbar^2c^2}{24m}n(1-n^2)N+ \frac{n\hbar^2}{2m}\sum_{j=1}^N k_j^2.
\end{equation}
In the last equation, we have used 
Eq. (\ref{lam_ansatz_3}). The Bethe ansatz Eq. (\ref{int_bethe_eq})
has been analyzed in Ref. \cite{Kardar87} to obtain the free energy of
the line lattice to leading order in the density $\rho=N/W$. The
central result of this reference is an integral equation for
integer $n$, which was obtained from Eq. (\ref{int_bethe_eq}) in the
limit $W,N \to \infty$ with $\rho$ fixed, by using the fact that
$[W(k_{j+1}-k_j)]^{-1}$ becomes a continuous function in that
limit. By taking the limit $n\to 0$, and subsequently the limit
$\rho\to 0$, the kernel of the integral equation becomes of Hilbert
type, thus enabling an analytic solution \cite{Tricomi57}.  To
obtain the probability distribution function of the free energy, we
have to determine the ground state energy as a function of $n$ for
small but finite arguments. Since extracting this information by an
exact method from the integer $n$ integral equation of
Ref. \cite{Kardar87} is not obvious, here we start
with a direct analytical continuation of the Bethe ansatz
Eq. (\ref{int_bethe_eq}) to non-integer $n$.

\subsection{Analytic continuation in $n$}

\begin{figure}
\begin{center}
  \resizebox{0.65\textwidth}{!}{ \includegraphics{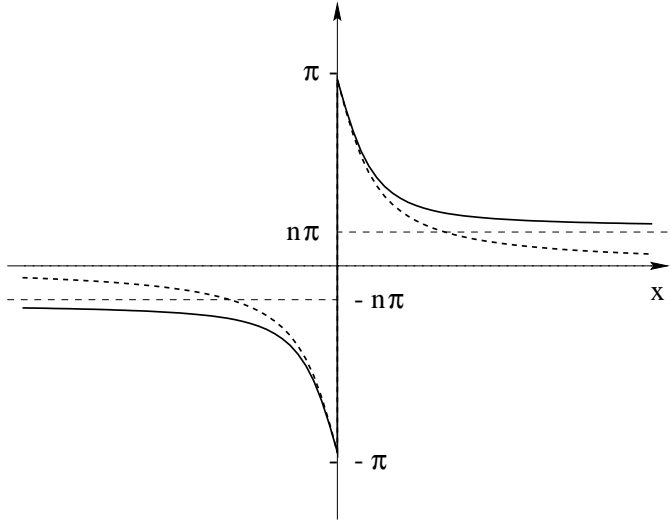} }
    \caption{Both the function $g_n(x)$ (solid curve) and its limiting
    form $2\arctan(n/x)=\pi\,\text{sgn}(x)-2\arctan(x/n)$ (dashed
    curve) for $x\ll 1$, $n \ll 1$ show a jump of $2\pi$ at $x=0$.}
    \label{fig1}
\end{center}
\end{figure}

In what follows, we show how to evaluate the right hand side (rhs) of
Eq. (\ref{int_bethe_eq}) for non-integer values of $n$. The product
over the replica index $\alpha$ represents simply the extension of a
factorial to complex numbers. Therefore, we can make use of the
recursion relation for the complex gamma function,
$\Gamma(z+1)=z\Gamma(z)$, to obtain for real valued $k$ the relation
\begin{eqnarray}
\prod_{\alpha=1}^{n-1}
\frac{ik/c+\alpha}{ik/c-\alpha}&=&(-1)^{n-1}
\frac{\Gamma(1-ik/c)\Gamma(n+ik/c)}{\Gamma(1+ik/c)\Gamma(n-ik/c)}\nonumber\\
&=&(-1)^n\frac{\Gamma(-ik/c)\Gamma(n+ik/c)}
{\Gamma(ik/c)\Gamma(n-ik/c)},
\end{eqnarray}
which is well defined for all $n$. Using this representation of the
finite product, and taking the logarithm of both sides of
Eq. (\ref{int_bethe_eq}), the proper extension of the Bethe ansatz
equations to real valued $n$ reads
\begin{equation}
\label{ba_real_n_1}
n k_j W = -i\sum_{\substack{l=1\\l\neq j}}^N \ln \left[
\frac{\Gamma(n+i(k_j-k_l)/c)\Gamma(-i(k_j-k_l)/c)}
{\Gamma(n-i(k_j-k_l)/c)\Gamma(i(k_j-k_l)/c)}
\right],
\end{equation}
where some care has to be taken to conserve the property that the
ground state has a total momentum of zero, i.e., $\sum_{j=1}^N k_j =
0$. To express the argument of the logarithm in terms of more
elementary functions, we rewrite it as the infinite product
\begin{equation}
\frac{\Gamma(n+ik/c)\Gamma(-ik/c)}{\Gamma(n-ik/c)\Gamma(ik/c)}
=\prod_{m=0}^\infty \frac{m^2+nm+(k/c)^2+ink/c}{m^2+nm+(k/c)^2-ink/c}.
\end{equation}
Here we have used the product representation $1/\Gamma(z)=ze^{\gamma
z} \prod_{m=1}^\infty (1+z/m)e^{-z/m}$ of the gamma function with
$\gamma$ the Euler constant \cite{Gradshteyn}. Next we write the last
expression as a product of exponential factors to compensate the
logarithm in Eq. (\ref{ba_real_n_1}). Using $\arctan(z)=
\frac{1}{2i}\ln((1+iz)/(1-iz))$ we obtain
\begin{equation}
\frac{\Gamma(n+ik/c)\Gamma(-ik/c)}{\Gamma(n-ik/c)\Gamma(ik/c)}
=\prod_{m=0}^\infty \exp\left[
2i\arctan \left( \frac{nk/c}{m^2+nm+(k/c)^2}
\right)\right].
\end{equation}
This result shows that the rhs of Eq. (\ref{ba_real_n_1}) is indeed a
real valued expression, and the final Bethe ansatz equation for
arbitrary real $n$ becomes
\begin{equation}
\label{final_bethe}
n k_j W = \sum_{\substack{l=1\\l\neq j}}^N g_n\left(\frac{k_j-k_l}{c}
\right),
\end{equation}
with the function
\begin{equation}
\label{gn(x)}
g_n(x)=2\sum_{m=0}^\infty\arctan\left( \frac{nx}{m^2+nm+x^2} \right).
\end{equation}
This is the central result of this section. The function $g_n(x)$ has
no poles for real valued $x$, is bounded between $\pm\pi$ and
converges to $\pm n\pi$ for $x \to \pm\infty$, see
Fig. \ref{fig1}. Notice that the jump of $2\pi$ at $x=0$ is produced
solely by the $m=0$ term in Eq. (\ref{gn(x)}), whereas the remaining
terms are continuous.

This Bethe ansatz equation is valid for all $n$ and for all line
densities $\rho$. To compare it with results already existing in
literature we consider two limiting cases. For $n=1$ we have
$g_n(x)=\pi\,\text{sgn}(x)$, where $\text{sgn}(x)$ is the sign
function.  The solution of Eq. (\ref{final_bethe}) then simply
consists of momenta $k_j$ which are equally spaced between $-\pi\rho$
and $\pi\rho$ for $W$, $N\to\infty$. This is the free fermion
situation corresponding to lines which wander due to thermal
fluctuations only. Notice that with only one color of lines ($n=1$)
present, the disorder induced inter-color interaction is, of course,
inactive. The same situation is expected to appear in the zero
disorder limit $c\to 0$. Indeed, in this limit $g_n(x\sim
1/c\to\infty)\to n\pi\,\text{sgn}(x)$ and the $n$'s on both sides of
Eq. (\ref{final_bethe}) cancel, leading again to the free fermion
result. The solution of Eq. (\ref{final_bethe}) in the limit $n\to 0$
gives the disorder averaged free energy of lines in a random
environment following Eqs. (\ref{Fcumulants}) and
(\ref{gsenergy}). Expanding $g_n(x)$ to first order in $n$, we get
from Eq. (\ref{final_bethe})
\begin{equation}
\label{bethe_coth}
k_jW=\sum_{\substack{l=1\\l\neq j}}^N \left[ \frac{c}{k_j-k_l} +
\pi\coth\left(\frac{\pi(k_j-k_l)}{c}\right)\right],
\end{equation}
where we have used the representation $\pi \coth(\pi x)=1/x+ 2x
\sum_{m=1}^\infty (m^2+x^2)^{-1}$ of the hyperbolic cotangent. It is
easily seen that this equation corresponds to the discrete momenta
version of the integral equation derived by a different technique for
treating the $n\to 0$ limit in Ref. \cite{Kardar87}. In this reference
the corresponding integral equation is solved in the dilute limit
$\rho/c\to 0$ yielding a free energy with steric interaction between
lines proportional to $\rho^2$ (instead of the behavior $\sim \rho^3$
due to thermal fluctuations). From Eq. (\ref{bethe_coth}), 
the free energy can be obtained at {\em all} densities \cite{Emig01-1}. 

Now consider the thermodynamic limit $N$, $W\to\infty$ of
Eq. (\ref{final_bethe}), while keeping the density $\rho=W/N$ fixed.
The employed technique is analogous to that of Lieb and Liniger
\cite{Lieb63}, using the fact that the function
$\varrho(k_j)=[W(k_{j+1}-k_j)]^{-1}$ becomes continuous in this limit.
The limiting behavior gives the density of states in the sense that
$W\varrho(k)dk$ = number of allowed $k$-values in the interval
[k,k+dk].  For a given density $\rho$, the boundary (Fermi) momentum
$K\equiv k_N =-k_1$ is determined by the condition
\begin{equation}
\label{def_K}
\int_{-K}^K \varrho(k) dk = \rho,
\end{equation}
while the ground state energy is given by
\begin{equation}
\label{e0_int}
E_0= \frac{\hbar^2c^2}{24m}n(1-n^2)\rho W+
\frac{\hbar^2nW}{2m}\int_{-K}^K k^2 \varrho(k) dk.
\end{equation}
Calculating the differences of Eq. (\ref{final_bethe}) between all
adjacent momenta $k_{j+1}$ and $k_j$ and expanding with respect to the
difference $k_{j+1}-k_j$, yields an integral equation for $\varrho(k)$,
\begin{equation}
\label{gen_inteq}
nk=\int_{-K}^K g_n\left(\frac{k-k'}{c}\right) \varrho(k') dk',
\end{equation}
for all $k\in [-K,K]$. This integral equation is not amenable to a
closed form solution because of the complicated form of its kernel
$g_n(x)$. However, the kernel is non-singular, and thus a numerical
treatment seems feasible, which we leave for a future publication
\cite{Emig01-1}. Instead, in the next Section we will show that taking
the limit of small $\rho/c$ leads to an interesting simplification
which paves the way for a description of the replicated lines in terms
of a pure, weakly interacting Bose gas.

\section{Mapping to a weakly repulsive Bose gas}

In this section, we develop a novel mapping of the dilute disordered
line lattice onto a weakly interacting pure Bose gas. This mapping
then provides a valuable tool to obtain exact results for the
probability distribution of the free energy in the important limits of
low density $\rho$ or high disorder strength $\Delta \sim c$. This is
the relevant limit for the critical behavior at the transition to a
line-free system, e.g., for flux lines expelled at $H_{c1}$ from
superconductors or for domain walls at the commensurate-incommensurate
transitions in absorbed layers or charge density waves.

From Eq. (\ref{def_K}) one observes that the cutoff momentum $K$ goes
to zero in the limit of $\rho\to 0$. Therefore the behavior of $g_n(x)$
for $x\ll 1$ determines the solution of the integral
Eq. (\ref{gen_inteq}) in the limit $\rho/c \ll 1$. This limit has
to be analyzed very carefully since we have to keep the $n$ dependence
of $g_n(x)$ to all orders to obtain a complete
expansion of the ground state energy $E_0$ in $n$, which is exact to
first order in $\rho/c \ll 1$.  Expressing the arctan of the $m=0$
part of Eq. (\ref{gn(x)}) in terms of its inverse argument, and
expanding the remaining terms in $x$, we obtain
\begin{equation}
\label{gn(x)_exp}
g_n(x)= \pi\,\text{sgn}(x)-2\arctan\left(\frac{x}{n}\right)+ 
x \sum_{m=1}^\infty \frac{2n}{m(m+n)} + {\mathcal O}(x^3).
\end{equation}
Note that we are interested in an expansion of the ground state energy
$E_0$ around $n=0$. Independent on how small the ratio $\rho/c$ and
therefore the argument of $g_n(x)$ is, the final expansion in $n$ for
fixed $\rho/c$ tests arbitrarily small values of $n$ and thus can render
the $\arctan(x/n)$ in Eq. (\ref{gn(x)_exp}) of order one. Therefore,
an expansion of this term is not justified.  In contrast, the
remaining terms with $m \ge 1$ in Eq. (\ref{gn(x)_exp}) tend always to
zero for both $x\ll 1$ and $n \ll 1$ and can be safely neglected.
Thus, retaining the first two terms of Eq. (\ref{gn(x)_exp}) only [see
Fig. \ref{fig1}] and taking the derivative of Eq. (\ref{gen_inteq})
with respect to $k$, we obtain the integral equation
\begin{equation}
\label{LL_inteq_o}
\int_{-K}^K\frac{2nc}{(nc)^2+(k-k')^2}\,\varrho(k')\,dk'=
2\pi\varrho(k)-n,
\end{equation}
which yields the complete $n$ dependence of $\varrho(k)$ to first
order in $\rho/c$ exactly. The crucial observation is that this
equation is identical to that obtained by Lieb and
Liniger for their exact Bethe ansatz solution of the one-dimensional
Bose gas with repulsive delta-function interactions \cite{Lieb63}. 

To obtain a physical understanding of the relation between the
replicated line lattice and the Bose gas, we change variables as
follows: Define the function $\varrho_b(k)=\varrho(k/n)/n$, the cutoff
momentum $K_b=nK$, the mass $m_b=nm$ and the rescaled interaction
strength $c_b=n^2c$. With this new variables Eqs.  (\ref{LL_inteq_o}),
(\ref{def_K}) read
\begin{equation}
\label{Lieb_inteq}
\int_{-K_b}^{K_b}\frac{2c_b}{c_b^2+(k-k')^2}\,\varrho_b(k')\,dk' =
2\pi\varrho_b(k)-1,\quad
\int_{-K_b}^{K_b} \varrho_b(k)dk = \rho.
\end{equation}
The bosonic ground state energy, i.e., the second term of
Eq. (\ref{e0_int}) becomes
\begin{equation}
\label{e0b}
E_{0,b}=\frac{\hbar^2W}{2m_b}\int_{-K_b}^{K_b} k^2 \varrho_b(k) dk.
\end{equation}
These equations represent the exact ground state (with energy $E_{0,b}$)
of a Bose gas with {\em repulsive} interactions, described by the
Hamiltonian
\begin{equation}
\label{bosegas_ham}
\hat H_b =-\frac{\hbar^2}{2m_b}\sum_{i=1}^N
\frac{\partial^2}{\partial x_i^2} +2c_{0,b}\sum_{i<j}
\delta(x_i-x_j),
\end{equation}
for $N\to \infty$ with fixed density $\rho$ of bosons
\cite{Lieb63}. Due to the definition of $c=2mc_0/\hbar^2$ the
amplitude of the interaction is given by $c_{0,b}=nc_0$. The
correspondence can be summarized as follows: The exact ground state
solution of the quantum Hamiltonian in Eq. (\ref{bethe_ham}) for {\em
attracting} particles, each appearing in $n$ colors, is in the dilute
limit $\rho/c \ll 1$, and to all orders in $n$ around $n=0$, the same
as the ground state solution of the Bose gas Hamiltonian of
Eq. (\ref{bosegas_ham}) with {\em repulsive} interactions. For
comparison with the usual mapping between elastic lines and world
lines of bosons for pure systems, both the correspondence of
parameters for the pure case and the novel analogy of quantities in
the random system are summarized in Tables \ref{tab1} and \ref{tab2}.

\begin{table}[h]
\caption{\label{tab1} The usual correspondence of the parameters of a
$d-1$-dimensional Bose gas and a $d$-dimensional lattice of lines
without a random potential. The interaction potential between vortex
lines maps to the Boson pair potential. In $d=2$ dimensions,
non-crossing lines are modeled by hard core bosons or, equivalently,
free fermions.}
\vspace{0.3cm}
\begin{center}
\begin{tabular}{|c|ccc|}\hline\hline
model & mass & Planck's & inverse \\
& & constant & temperature \\\hline
Bose gas & $m_b$ & $\hbar$ & $\hbar\beta_{\rm qm}$\\\hline
non-random lines & $g$ & $T$ & $L$\\ 
\hline\hline
\end{tabular}
\end{center}
\end{table}

\begin{table}[h]
\caption{\label{tab2}The correspondence of parameters in the novel
analogy between the one-dimensional Bose gas and the two-dimensional
replica theory of lines. The density of bosons and (non-replicated)
lines, respectively, is in both descriptions given by $\rho$.}
\vspace{0.3cm}
\begin{center}
\begin{tabular}{|c|ccccc|}\hline\hline
model & mass & Planck's & inverse & contact  & effective \\
& & constant & temperature& interaction & interaction \\\hline
Bose gas & $m_b$ & $\hbar$ & $\hbar\beta_{\rm qm}$ &$c_{0,b}$ & $c_b$ \\ 
& & & & (repulsive) & $\equiv 2m_bc_{0,b}/\hbar^2$ \\ \hline
replicated & $nm$ & $T$ & $L$ & $nc_0\equiv n\Delta/2T$ & 
$n^2c$ \\ 
lines & $\equiv ng$ & & &(attractive) & $\equiv n^2g\Delta/T^3$ \\ 
\hline\hline
\end{tabular}
\end{center}
\end{table}

\begin{figure}
\begin{center}
  \resizebox{0.75\textwidth}{!}{ \includegraphics{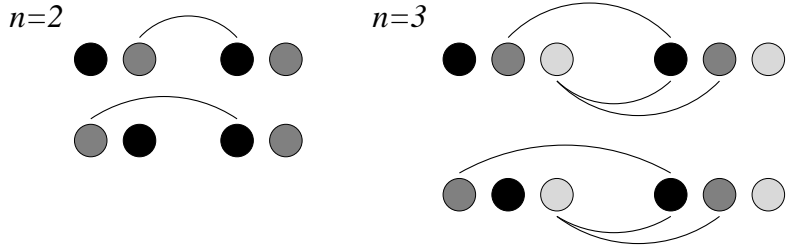} }
    \caption{The origin of the effective cluster-cluster interaction:
    For $n=2$ there is for both configurations only one inter-cluster
    interaction possible since the black particle of the first cluster
    is blocked by the black particle of the second cluster from
    interacting with the dark gray particle. For $n=3$ only two
    typical configurations are shown. It is easy to convince oneself
    that each of the $3!$ configurations allows for three interactions
    suggesting $1+\ldots+(n-1)=\frac{1}{2}n(n-1)$ pair interactions
    for $n$-particle clusters. \label{fig:clusters}}
\end{center}
\end{figure}

The correspondence between the Bose gas and the replica theory of
lines can be interpreted physically as follows: As we have seen in
Section \ref{sec:bethe}, the Bethe ansatz solution can be interpreted
as $n$ lines of different colors forming a cluster due to the pair
attraction of amplitude $c_0$. This suggests considering these
clusters as new particles of mass $nm=ng$, then appearing as the
components of the Bose gas described by Eq. (\ref{bosegas_ham}).  To
obtain insight in the effective strength of the contact interaction
between these composite bosons, consider two clusters, each composed
of $n$ lines of different colors.  Since particles of the same
color avoid each other like free fermions or, equivalently, bosons
with hard core repulsion, some inter-cluster interactions are
``screened,'' see Fig. \ref{fig:clusters}. There are $n!$ orderings of
particles within a cluster, each ordering providing the same number of
pair interactions given by $1+\ldots+(n-1)=\frac{1}{2}n(n-1)$ due to
screening. Finally, we have to take into account that in the quantum
Hamiltonian (\ref{bethe_ham}), which corresponds to the replica theory
of lines, the interaction sum runs implicitly for two given clusters
over all different colors $\alpha\neq\beta$ instead of running over
color pairs $\alpha<\beta$ only.  Therefore, the total amplitude of
the contact interaction between two clusters is given by $n(n-1)c_0$,
which yields a repulsion of amplitude $-nc_0$ in the limit $n\ll 1$,
in agreement with our findings (see Table \ref{tab2}). That this naive
estimate produces the correct result to first order in $\rho/c$
is presumably related to the fact that the internal structure of the
clusters does not matter in the dilute limit since the distance
between clusters is much larger than cluster size $\sim 1/c$.

Having established the analogy between the replica theory of lines and
the Bose gas, one can take advantage of the many results for the
one-dimensional Bose gas accumulated during the past
decades \cite{Lieb63,Takahashi75,Haldane81,Korepin97}. In the
following, we focus on the ground state energy $E_{0,b}$ as given by
Eq. (\ref{e0b}) to obtain the free energy $F_n$ of the replicated line
system using Eq. (\ref{F_n}). The only dimensionless intensive
variable in the Bose gas problem is $\gamma=c_b/\rho=n^2g\Delta/\rho
T^3$. Thus the ground state energy can be written as 
\begin{equation}
\label{def_e}
E_{0,b}=\frac{\hbar^2}{2m_b}W\rho^3 e(\gamma),
\end{equation}
where $e(\gamma)$ is a dimensionless monotonically increasing function
of $\gamma$ \cite{Lieb63}. Since we are interested in the limit $n\ll
1$, we have to consider the weak coupling limit of the Bose gas. In the
limit $c_b\to 0$, the integral equation (\ref{Lieb_inteq}) has the
diverging\footnote{The divergence is related to the fact that as
$c_b\to 0$ the kernel of the integral equation is a representation for
$2\pi\delta(k)$ leading to $2\pi\varrho_b(k)-1\to
2\pi\varrho_b(k)$. Therefore, $\varrho_b(k)$ has to grow unbounded in
this limit.} solution \cite{Lieb63}
\begin{equation}
\label{varrho_b_sol}
\varrho_b(k)=\frac{1}{2\pi c_b}\sqrt{K_b^2-k^2}.
\end{equation}
This zero-order result corresponds to the $n=0$ limit of $\varrho(k)$,
which is a well-defined expression according to the mapping between
the two models, and given by
\begin{equation}
\lim_{n\to 0}\varrho(k)=\frac{1}{2\pi}\frac{T^3}{g\Delta}\sqrt{K^2-k^2},
\end{equation}
with $K=2\sqrt{g\Delta\rho/T^3}$.  This result coincides with that 
obtained in Ref. \cite{Kardar87} by solving an integral equation which
itself is valid for $n=0$ only. Using Eqs. (\ref{Lieb_inteq}),
(\ref{e0b}) one gets $K_b=2\rho\gamma^{1/2}$, $e(\gamma)=\gamma$ and,
therefore, the non-trivial (disorder dependent) part of the disorder
averaged free energy of the lines is given by
\begin{equation}
\label{free-energy}
[F]_d=\lim_{n\to 0}\frac{E_0(n)}{n}L=\frac{\Delta}{2T}\left(
\frac{1}{12}\frac{g\Delta}{T^3}\rho+\rho^2\right)WL,
\end{equation}
using Eqs. (\ref{Fcumulants}), (\ref{e0_int}), (\ref{def_e}), and the
correspondences of parameters in Table~\ref{tab2}. Note the difference
from the case without random potential, in which the non-linear term
in $[F]$ is proportional to $\rho^3$.

\section{Distribution of the free energy in the dilute limit}

The preceding Section demonstrates how the lowest order term in the
energy of the Bose gas yields the first moment or cumulant of the free
energy of the random line system in the limit $\rho\to 0$.  As
discussed in connection with a scaling hypothesis for PDF's of
disordered system \cite{Aharony96,Emig00}, cumulants of the free
energy are also singular at critical points.  To calculate these
cumulants the complete $n$ dependence of the replica free energy $F_n$
is needed. Due to the mapping between replicated lines and bosons,
$F_n$ can be obtained order by order around $n=0$ by perturbation
theory for the Bose gas in the weak coupling limit $c_b=n^2
g\Delta/T^3 \ll 1$. As realized by Lieb and Liniger \cite{Lieb63}, and
as the zero-order solution [see Eq. (\ref{varrho_b_sol})] suggests,
the complete solution of the integral equation (\ref{Lieb_inteq}) is
highly singular at $c_b=0$. The physical meaning of this singularity
in the context of the line system is related to the relevance of the
random potential. Any randomness, however weak, leads to a glassy
state that is fundamentally different from the pure case.

The singular behavior of the integral equation seems to preclude any
direct analytical way of extracting higher order corrections to the
zero-order result.  Fortunately, for small $\gamma$, Bogoliubov's
perturbation theory \cite{Bogoliubov47} can be used to calculate the
ground state energy of the Bose gas.  One reason for trusting
Bogoliubov's theory for small $\gamma$ is that in the case of a
delta-function potential the ground state energy is given correctly by
this theory in the limit of {\em high} boson density $\rho$
\cite{Lieb63}. Note that the important small $n$ limit maps onto the
high density limit of the Bose gas as far as the coupling strength
$\gamma \sim n^2/\rho$ is concerned, although we are studying the low
density limit of the line system. Indeed, agreement between the exact
ground state energy obtained numerically from the integral equation
(\ref{Lieb_inteq}) and Bogoliubov's first-order perturbation theory
has been confirmed for small $\gamma$ by Lieb and Liniger
\cite{Lieb63}. Later Takahashi \cite{Takahashi75} used the correlated
basis function method to calculate the second-order correction to the
ground state energy, which again coincides with the second-order
energy in Bogoliubov's perturbation theory \cite{Lee71}, and is in
perfect agreement with a numerical high-accuracy solution of the
integral equation (\ref{Lieb_inteq}) down to $\gamma \sim 10^{-5}$
\cite{Emig01-2}. Using the analytical results for first and second
order terms, the next two orders can be obtained by this numerical
solution, and the ground state energy summarized, in terms of the
rescaled function introduced in Eq. (\ref{def_e}), as
\begin{equation}
\label{e_res}
e(\gamma)=\gamma-\frac{4}{3\pi}\gamma^{3/2}+\left(\frac{1}{6}-
\frac{1}{\pi^2}\right)\gamma^2+b_5 \gamma^{5/2}+b_6 \gamma^3 +
{\mathcal O}(\gamma^{7/2}),
\end{equation}
with numerical coefficients \cite{Emig01-2}
\begin{equation}
b_5=-0.001588,\quad b_6=-0.000171.
\end{equation}
It is interesting to note that the ground state energy is non-analytic
in the coupling strength and thus has to be considered as an
asymptotic expansion for small $\gamma$.

The replica free energy $F_n$ of the line system can now be evaluated
by applying the mapping between bosons and lines. First, we note that
Eq. (\ref{e_res}) provides an explicit proof that $F_n$ is indeed an
analytic function of $n$ since $\gamma\sim n^2$. While this property
of $F_n$ is usually assumed in replica theories, it cannot be proved
in most cases, and had not been checked before for random line
lattices. In terms of the cluster size
\begin{equation}
\label{l_d}
l_d=\frac{1}{c}=\frac{T^3}{g\Delta}, 
\end{equation}
which sets the crossover length scale from pure to disorder dominated
behavior for a single line \cite{Kardar87}, the disorder dependent
part of the replica free energy can be written as
\begin{equation}\begin{split}
F_{n,d}&=\frac{LW}{l_d^2}\frac{\Delta}{2T}\bigg\{ \frac{n}{12}(1-n^2)l_d 
\rho +n (l_d\rho)^2-\frac{4}{3\pi}n^2 (l_d\rho)^{3/2}\\ 
&\quad + \left(\frac{1}{6} -\frac{1}{\pi^2}\right)n^3 l_d\rho + b_5 n^4 
(l_d\rho)^{1/2} + b_6 n^5 + {\mathcal O}(n^6)\bigg\},
\end{split}\end{equation}
in the dilute limit $\rho \ll l_d^{-1}$. This exact result for the
replica free energy has the scaling form
\begin{equation}
\label{F_scal}
F_n=LWn\frac{\Delta\rho^2}{T}\,{\mathcal G}\left(n(l_d\rho)^{-1/2}\right),
\end{equation}
where ${\mathcal G}(x)$ is a polynomial. This result is in agreement with
dimensional arguments given in Ref. \cite{Kardar87}, which motivated a
scaling theory for cumulants of thermodynamic quantities in random
systems \cite{Emig00}. Thus this scaling theory is strikingly
confirmed by our exact results.

The cumulants of the free energy of the line lattice at fixed density
$\rho$ can be read off from $F_n$. They can be written in the general
form
\begin{equation}
[F^p]_c=\sigma_p \frac{WL}{l_d^2}\Delta T^{p-2} (l_d \rho)^{(5-p)/2}
\end{equation}
with coefficients
\begin{equation}
\sigma_1=\frac{1}{2},\quad \sigma_2=\frac{4}{3\pi}, 
\quad \sigma_3=\frac{1}{4}-\frac{3}{\pi^2}, \quad \sigma_4=0.01906,
\quad \sigma_5=-0.01026. 
\end{equation}
Consistent with the central limit theorem all cumulants of the free
energy {\em density} vanish in the thermodynamic limit as $[f^p]_c
\sim (LW)^{1-p}$. However, for a large but finite system the cumulants
are finite and show a non-trivial dependence on the density, or on
the chemical potential $\mu$ within a grand canonical
description. Close to the critical point we have $\rho \sim \mu-\mu_c$
and $[f^p]_c \sim (\mu-\mu_c)^{(5-p)/2}$. Therefore, as $\mu\to\mu_c+$
the first four cumulants vanish continuously at the critical point,
the fifth cumulant approaches a constant and higher order cumulants
show increasing divergence\footnote{A numerical solution of the
integral equation (\ref{Lieb_inteq}) indicates that the coefficient
$b_7$ in Eq.  (\ref{e_res}) is negative and about one order of
magnitude smaller than $b_6$ \cite{Emig01-2}. This suggests that the
sixth and presumably all higher cumulants have indeed non-vanishing
contributions $\sim WL$.  }.  

Of course, since we are discussing finite systems, the singular
behavior is cutoff when the correlation length approaches the system
size. Due to the anisotropy of the line system, there are two
correlation lengths. The first one is given by the mean distance
between lines, $\xi_\perp=1/\rho$, whereas the second one is set by
the mean longitudinal distance between collisions of lines,
$\xi_\|=(gT/\Delta)^{1/2} \xi_\perp^{3/2}$. The critical regime
corresponds to $\xi_\perp \gg W$, $\xi_\| \gg L$. We consider a fixed
ratio of longitudinal and transversal system sizes so that they are in
agreement we the anisotropic scaling of the system, i.e.,
$L=\delta\xi_\| (W/\xi_\perp)^{3/2}$, allowing for an anisotropy parameter
$\delta$. In the critical regime, the moments of the free energy
density are then given by $[f^p]_c=\sigma_p (\Delta/TW^2)^p$ yielding
{\em universal} relative cumulants
\begin{equation}
\frac{[f^p]_c}{[f]^p}=2^p\sigma_p\delta^{1-p},
\end{equation} 
which depend only on the geometry of the system via the anisotropy
parameter $\delta$. This result is a consequence of the relevance of
disorder, and demonstrates the complete destruction of self-averaging
at the critical point, since otherwise the relative cumulants vanish
with increasing system size \cite{Aharony96}. Numerical simulations
for systems of appropriate size ratios should be suitable to test this
universality. Another interesting property of the PDF of the free
energy is its asymmetry. Since the third and fifth cumulant are both
negative, there is a higher probability for the system to be in a
state with a free energy which is {\em smaller} than the average
value. This asymmetry gets more pronounced as the system gets closer
to the critical point.

Some comments on the above results for the PDF are in order. These
comments are of more general nature, and are given here in order to
clarify the status of the replica method itself. The replica approach
provides a tool to calculate integer moments $[Z^n]$ of the partition
function. However, $Z$ itself is unphysical, and the desired PDF of
$\ln Z$ has to be deduced using analytical continuation of the moments
$[Z^n]$ to non-integer $n$, and applying the cumulant series
representation in Eq. (\ref{Fcumulants}). There are two potential
problems related to this kind of procedure. First, an unique PDF can
be deduced from the integer moments of the PDF only if the $[Z^n]$
grow slower than $n!$ for {\it large} $n$ \cite{Akhiezer65}. One might
conclude that this condition is not fulfilled for the random line
system because of $[Z^n]\sim \exp(-E_0(n) L/T)$ and $E_0(n)$ has a
contribution proportional to $n(1-n^2)$, see Eq. (\ref{e0_int}). The
origin of this contribution can be traced back to the energy of
individual clusters in the Bethe ansatz. Therefore, the same situation
appears for the single line system. In this context, it has been noted
that the result for $E_0(n)$ given above is not expected to be valid
for arbitrarily large $n$ \cite{Kardar96}. One is lead to this
conclusion by the observation that the cluster size goes to zero if
$n$ tends to infinity, reflecting the tendency of {\em attractive}
bosons to collapse. The issue of an attracting 1D Bose gas has been
discussed briefly by Lieb and Liniger \cite{Lieb63} in connection with
their Bethe ansatz solution. They conclude that for the attractive
Bose gas it is not clear that the Bethe ansatz wavefunction which
includes the $n$-string solution exhausts all solutions of the
Schr\"odinger equation. When the cluster size $l_d$ goes to zero, it
seems to be no longer justified to construct the wavefunction as a
product of only two-paricle functions $\exp(-|x_1-x_2|/l_d)$, as done
for the $n$-string, since they are extremely localized.  The second
potential difficulty of the replica method has its origin in the order
in which the thermodynamic limit and the $n \to 0$ limit are taken. To
apply the cumulant series in Eq. (\ref{Fcumulants}), we must expand
the analytically continued function $\ln [Z^n]$ around $n=0$ for a
{\it fixed} system size. However, in calculating $[Z^n]$ in terms of
the quantum system, first we consider at {\em fixed} $n$ the
thermodynamic limit to eliminate all exicted states. For a single line
in random media the two limits do not commute due to the non-trivial
sample-to-sample variation of the free energy \cite{Kardar96}, and the
free energy cumulants have to be obtained indirectly by a scaling
ansatz for the replica free energy \cite{Emig00}. For the line lattice
considered here, the situation is different. Now, the extensive
replica free energy $F_n$, obtained by first taking the thermodynamic
limit and then the limit of small $n$, has contributions at any order
in $n$, and thus the two limits are more likely to commute.

Finally, we would like to mention that our mapping between bosons and
elastic lines provides also new information about the weak coupling
limit of the ground state energy of the one-dimensional Bose gas.
Since cumulants $[F^p]_c$ of even order $p$ have to be positive, all
coefficients of non-integer powers of $\gamma$ in the expansion of the
ground state energy in Eq. (\ref{e_res}) have to be negative, see
Eq. (\ref{Fcumulants}). This is an important constraint on the various
approximative methods in use for the Bose gas. Recently, different
self-consistent approximations have been used to calculate the ground
state energy based on a so-called local-field correction approach
\cite{Gold93,Demirel99}. In view of the fact that the two approaches
give different signs for the $\gamma^{7/2}$ term in Eq. (\ref{e_res}),
our sign constraint rules out the solution of Ref. \cite{Gold93} from
being correct.

\section{Correlation functions}

The preceding study of the free energy of the line lattice shows that
disorder induces critical behavior which is distinct from the pure
case. Moreover, disorder is expected to lead to correlations in
the system which are different from pure 2D line lattices. In the
context of flux lines in high-$T_c$ superconductors, Fisher predicted a
``vortex glass'' phase in which disorder locks the flux lines into
one of many possible metastable states \cite{MPAFisher89}.  He showed
that the random line lattice can be mapped onto the two-dimensional
random-field XY (RFXY) model with vortices excluded.  The correlation
functions of the low temperature phase of the RFXY model, and the
vortex glass phase, and their collective excitations have been studied
extensively during the last decade. For the RFXY model, analytical
approaches include replica-symmetric renormalization group (RG)
techniques \cite{Cardy82,Goldschmidt82,Paczuski91}, and one-step
replica-symmetry breaking variational ans\"atze \cite{Korshunov93} and
RG methods \cite{LeDoussal94,Kierfeld95}.  The analytical studies of
the random line lattice are mainly based on RFXY model results. They
consist in scaling arguments \cite{Fisher91}, linear continuum
elasticity models \cite{Nattermann91}, RG techniques
\cite{Toner91,Nattermann92,Giamarchi95}, and variational methods
including replica-symmetry breaking \cite{Giamarchi95} and without use
of replicas \cite{Orland95}. However, the results obtained within the
different approaches are not consistent. Whereas all variational
methods predict for the continuum displacement field $u(x,y)$ of the
lines, correlations of the form $[\langle (u(x,y)-u(0,0))^2 \rangle]
\sim \ln(x^2+y^2)$, RG approaches yield additional corrections growing
as $\ln^2(x^2+y^2)$, which vanish as the non-glassy high-temperature
phase is approached. A critical comparison of the various results is
given in Ref. \cite{Nattermann00}. One should also keep in mind the
range of validity of the different approaches. The variational methods
are expected to hold throughout the whole glassy phase, while the RG
results are applicable only close to the glass transition point.

Numerical simulations of glassy line lattices in two dimensions are
also inconclusive.  For both the variational method predictions
\cite{Batrouni94,Cule95}, and the RG results
\cite{Lancaster95,Marinari95,Zeng96,Rieger97}, qualitative agreement
was found in simulations. The numerical calculations are either
hampered by slow glassy dynamics, and/or by a large length scale for
the crossover to a regime where differences between the two analytical
methods become significant. More recently, Zeng {\em et al.}
\cite{Zeng99} found good quantitative agreement, albeit in the
vicinity of the glass transition, between the RG results and their
numerical studies of the correlation functions via a mapping to a
discrete dimer model with quenched disorder.

The analogy between the line lattice and the quantum model of $U(n)$
fermions has been used as an alternative to RG for obtaining
information about correlations in the glassy phase
\cite{Tsvelik92,Balents93}. References \cite{Tsvelik92} and
\cite{Balents93} both use the replica-symmetric Bethe ansatz, but
employ different techniques in handling the $n\to 0$ limit, and make
different assumptions. As a consequence, the two approaches yield
different results for the asymptotic decay of the line lattice
density-density correlation function, i.e.,
$[\langle\rho(x,y)\rho(0,0)\rangle]-\rho^2\sim (x^2+y^2)^{-1}$
\cite{Tsvelik92} compared to
$[\langle\rho(x,y)\rho(0,0)\rangle]-\rho^2\sim (x^2+y^2)^{-1/2}$
\cite{Balents93}. Although the decay exponent of the first result is
in agreement with the variational ans\"atze, the amplitude calculated
in Ref. \cite{Tsvelik92} does not agree with them. Neither result
shows the logarithmic corrections predicted by the RG approaches.

In this section we reconsider the displacement and density
correlations in the glassy phase by exploring the mapping of the
random line lattice to the pure Bose gas. While the free energy of the
former is related to the ground state energy of the latter,
correlations in the statistical system are related to the low energy
excitations of the quantum problem.  Therefore, we calculate the
low-energy excitations of the $U(n)$ fermion model in the limit $n\to
0$. In the following, we consider only replica-symmetric or color-less
excitations, which do not break apart the $n$-particle clusters
corresponding to the bosons. This restriction is motivated by the fact
that for $n>1$ only singlet states are expected to be gap-less
\cite{Affleck87}, due to the finite energy associated with breaking up
a cluster. With this assumption, the excitations above the ground
state consist of ``cluster-particles'' and ``cluster-holes.''  Under
the mapping\footnote{That the mapping holds also for the excitation
spectrum follows from the fact that the response of the Fermi sea
(shifts of the discrete wave vectors $k_j$) to excitations is
completely determined by the phase shifts in an infinite system
associated with transposing two wave vectors of the Bethe wave
function, see, e.g., Ref. \cite{Lieb63}.} of the replicated lines or
$U(n)$ fermions to the Bose gas, these excitations become the usual
particle and hole states studied by Lieb \cite{Lieb63}. He calculated
the double spectrum associated with particle and hole elementary
excitations exactly by Bethe ansatz. Particle states can be excited
with any momentum $p$, but the hole state spectrum exists only for
momenta $p$ satisfying the condition $|p| \le \pi\rho$. Both spectra
have in common a {\em linear} dispersion relation at small momenta
with the same velocity of sound,
\begin{equation}
\epsilon_{\text{p,h}}(p)=v_s |p| + {\mathcal O}(p^2), 
\end{equation}
where $v_s$ in the weak coupling limit is given by 
\begin{equation}
v_s=\frac{\hbar
\rho \gamma^{1/2}}{m_b}=\sqrt{\frac{2\rho c_{0,b}}{m_b}}.
\end{equation}
Since under the mapping both $c_{0,b} \sim n$ and $m_b \sim n$, the
excitation spectra of the replicated lines or $U(n)$ fermions remain
linear in the $n \to 0$ limit, and the velocity of sound is given by
\begin{equation}
\label{mapped_vs}
\lim_{n\to 0} v_s  \to \sqrt{\frac{\rho\Delta}{gT}}. 
\end{equation}
A linear dispersion relation in the $n \to 0$ limit was also obtained
in Refs. \cite{Tsvelik92,Balents93} by solving directly a linear
integral equation which determines the shifts of the Bethe ansatz
momenta $k_j$ under particle/hole excitations. However, in these
references different assumptions and approximations were made to go
from the excitation spectra to the density-density correlation
function of the line lattice, leading to contradictory results.

Having the mapping to the pure Bose gas at hand, it is tempting to
reanalyze the density-density correlations of the line lattice from a
new perspective. Below we will show that the mapping gives direct
results only for certain correlations in the replicated system, from
which information about correlations in the original random system has
to be deduced. Therefore, the remaining part of the section is guided
more by the question of how much we can learn about the correlations
in the line system from the knowledge of the density correlation
function of the Bose gas, rather then giving final exact results for
the line lattice correlations.  The density-density correlation
function of the 1D Bose gas at zero temperature has been studied
extensively by several methods in the past, see, e.g.,
Ref. \cite{Korepin97}. Since the dispersion relation is linear for
small $p$, conformal field theory (CFT) can be employed to obtain the
asymptotics of the correlation function. The decay exponent $\eta_b$
is given by the conformal dimensions, which are determined by the
spectra of low-lying excitations.  For the class of Luttinger liquids,
including the Bose gas, the CFT approach is equivalent to Haldane's
effective harmonic fluid description of 1D quantum fluids
\cite{Haldane81}. This description is based on the fact that the
elementary excitations may be regarded as bosons which represent
long-wavelength density fluctuations. The Hamiltonian for these bosons
can be expressed as a sum of harmonic oscillators. Therefore,
correlation functions can be calculated by simple Gaussian
averages. In particular, from the Luttinger liquid approach and CFT,
the asymptotic time-dependent correlation function of the local
density $b(x,t)$ of the 1D Bose gas is known at any coupling strength
$c_b$ as \cite{Haldane81,Berkovich89}
\begin{equation}
\label{bose_corr_1}
\langle b(x,t)b(0,0)\rangle = \rho^2 - \frac{\eta_b}{4\pi^2}
\frac{x^2-v_s^2t^2}{(x^2+v_s^2 t^2)^2} + A \frac{\cos(2\pi\rho x)}
{(x^2+v_s^2t^2)^{\eta_b/2}}+\text{h.h.},
\end{equation}
where the coefficient $A$ is not known exactly and higher harmonics
(h.h.) are not shown explicitly. The coupling strength enters the
exponent $\eta_b$ only via the velocity of sound, which is known
perturbatively for small $c_b$, and numerically for all $c_b$, from
the Bethe ansatz \cite{Lieb63}. This exponent also determines the
amplitude of the homogeneous part of the correlations, and is given by
\begin{equation}
\eta_b=\frac{2\pi\hbar\rho}{m_b v_s}.
\end{equation}

We now examine to what extent the knowledge of the density-density
correlation functions of the Bose gas provides information about the
random line lattice correlations.  First consider the correlation
function of the replicated $U(n)$ fermion system of mean density
$n\rho$. Its local particle density $r(x,y)$ is given by the sum over
all densities $\rho_\alpha(x,y)$ of particles of a specific color
$\alpha$, i.e.,
\begin{equation}
r(x,y)=\sum_{\alpha=1}^n \rho_\alpha(x,y).
\end{equation}
Since all replicas or colors are equivalent, we obtain in terms of the
density fluctuations $\delta\rho_\alpha(x,y)=\rho_\alpha(x,y)-\rho$,
the relation
\begin{equation}
\langle r(x,y)r(0,0) \rangle = n^2\rho^2 + n  
\langle\delta\rho_\alpha(x,y)\delta\rho_\alpha(0,0)\rangle  
+n(n-1)\langle\delta\rho_\alpha(x,y)\delta\rho_\beta(0,0)\rangle .
\end{equation}
On length scales larger then the cluster size $l_d$, see
Eq. (\ref{l_d}), the simple relation $r(x,y)=nb(x,y)$ between boson
and $U(n)$ fermion densities holds, since bosons correspond to
$n$-clusters of fermions. Thus, we obtain
\begin{equation}\begin{split}
\label{bose_corr_2}
\langle b(x,y)b(0,0) \rangle & = \rho^2 + \frac{1}{n} \left(
\langle\delta\rho_\alpha(x,y)\delta\rho_\alpha(0,0)\rangle
-\langle\delta\rho_\alpha(x,y)\delta\rho_\beta(0,0)\rangle \right)\\
&\quad  +\langle\delta\rho_\alpha(x,y)\delta\rho_\beta(0,0)\rangle ,
\end{split}\end{equation}
where we have assumed that all off-diagonal correlations
($\alpha\neq\beta$) are identical.  The disorder and thermally
averaged density-density correlation function of the original line
system can be calculated by taking the $n \to 0$ limit of one of the
$n$ equivalent correlation functions for particles of the same color,
\begin{equation}
\label{def_ddc}
[\langle \rho(x,y)\rho(0,0)\rangle ]= \lim_{n\to 0} 
\langle \rho_\alpha(x,y)\rho_\alpha(0,0) \rangle .
\end{equation}
Can we extract information about this correlation function from the
Bose gas correlations?  Assuming that correlations of the density
fluctuations $\delta\rho_\alpha(x,y)$ have a finite limit for $n\to
0$, it follows from Eq. (\ref{bose_corr_2}) that we can expect two
types of terms in the $n\to 0$ limit of Eq. (\ref{bose_corr_1}):
Contributions diverging as $\sim 1/n$, and those which
saturate. Counting powers of $n$ shows that we cannot identify the
diagonal correlations needed in Eq. (\ref{def_ddc}), but only the
difference between diagonal and off-diagonal correlations. However,
this differene can be calculated both in the RG approach and the
variational ansatz (VA), and is therefore of interest, too.  To check
for the diverging contributions proportional to $1/n$ in
Eq. (\ref{bose_corr_1}), we have to map the Bose gas exponent $\eta_b$
to the corresponding exponent $\eta$ of the replicated line
lattice. Using the result of Eq. (\ref{mapped_vs}) for the mapped
velocity of sound, $\hbar \to T$ and $m_b \to ng$, we get in the limit
$n \to 0$ the diverging exponent
\begin{equation}
\label{eta}
\eta=\frac{2\pi}{n}\sqrt{\frac{\rho T^3}{g\Delta}}=
\frac{2\pi}{n}\sqrt{l_d \rho}.
\end{equation}
For the last expression, we have used the definition of the length
scale $l_d$ given in Eq. (\ref{l_d}).  Focusing on the homogeneous
part of the line density correlations, we obtain, by comparing the
diverging part proportional to $1/n$, the correlation function
\begin{equation}
\label{Bethe_corr}
[\langle \rho_\alpha(x,y)(\rho_\alpha(0,0)
-\rho_\beta(0,0))\rangle ]=- 
\frac{\rho^2}{2\pi}\sqrt{l_d \rho}\,\frac{(x/a)^2-(y/\xi_\|)^2}
{((x/a)^2+(y/\xi_\|)^2)^2}+ \text{o.t.},
\end{equation}
where o.t. stands for omitted oscillating terms. The length scales in
the last expression are $a=1/\rho$, the mean separation of lines, and
$\xi_\|=\sqrt{gT/\rho^3\Delta}$ corresponding to $1/\rho v_s$, the
mean longitudinal distance between collisions of lines.

Before we discuss the oscillating contributions of the correlation
function, we would like to compare our result for the difference
correlations in Eq. (\ref{Bethe_corr}) with the predictions made by RG
techniques and VA. Both, RG and VA are constructed on the basis of
linear elasticity theory, which describes the line lattice in terms of
a scalar displacement field $u(x,y)$. To represent the coupling of the
random potential to the line density in terms of the displacement
field, the line density in each replica $\alpha$ has to be decomposed
into harmonics as \cite{Haldane81,Nattermann90,Nattermann91}
\begin{equation}
\label{density_decomp}
\rho_\alpha(x,y)=\rho\left\{1-\partial_x u_\alpha(x,y)\right\}
\sum_{m=0}^\infty
\cos\{2\pi m\rho(x-u_\alpha(x,y))\}.
\end{equation}
In this representation, higher order terms in $\partial_x
u_\alpha(x,y)$ are usually ignored since $u_\alpha(x,y)$ varies slowly
on coarse-grained scales. By retaining only the coupling of the first
harmonic ($m=1$) to the random potential, one obtains for the
displacement field $u(x,y)$ the Hamiltonian of the RFXY model
\cite{MPAFisher89}. For this model, the RG approach
\cite{Goldschmidt82} gives the result
\begin{eqnarray}
\label{RG_corr}
[\langle\partial_x u_\alpha(x,y)\partial_x
u_\alpha(0,0)\rangle] &=& \frac{\partial^2}{\partial x^2}
\frac{a^2}{4\pi^2}\left\{
\frac{T}{2\pi J} \ln \left(r/a\right)
+ \tau^2 \ln^2  \left(r/a\right) \right\},\\
\label{RG_corr_2}
{}[\langle\partial_x u_\alpha(x,y)\partial_x u_\beta(0,0)\rangle ] 
&=& 
\frac{\partial^2}{\partial x^2}
\frac{a^2}{4\pi^2}\tau^2 \ln^2  \left(r/a\right),
\end{eqnarray}
where the $\ln^2$ contribution is obtained by an expansion to lowest
order in $\tau=1-T/T_g$ which measures the distance from the glass
transition at $T_g=4\pi J$. Here $J$ is the elastic constant of the
(isotropic) RFXY model and $r^2=x^2+(c_{11}/c_{44})y^2$ the rescaled
squared distance. $J$ can be expressed in terms of the elastic
constant $c_{11}$ and $c_{44}$, which are related to compressions and
tilts of the line lattice, by \cite{Nattermann91,Nattermann00}
\begin{equation}
J=\frac{a^2}{4\pi^2}\sqrt{c_{11}c_{44}}.
\end{equation}
In contrast, VA \cite{Giamarchi95} yields a simple logarithmic growth
of displacements with $\tau=0$. The amplitude of the logarithm in
Eq. (\ref{RG_corr}) then depends on whether one allows for replica
symmetry breaking or not. The replica symmetric result for the
amplitude agrees with the RG result in Eq. (\ref{RG_corr}), as
compared to the universal amplitude $a^2/2\pi^2$, which is obtained in
case of replica symmetry breaking.

The relation between density-density correlations and displacement
correlations follows from Eq. (\ref{density_decomp}) as
\begin{equation}
\label{d-u-rel}
[\langle \rho_\alpha(x,y)\rho_\beta(0,0) \rangle] = \rho^2 \left( 1 +
[\langle \partial_x u_\alpha(x,y) \partial_x u_\beta(0,0) \rangle] \right) +
\text{o.t.},
\end{equation}
where we have not written explicitly the oscillatory contributions
which are generated by harmonics of order $m>1$ in the density
decomposition. Using this relation, we can extract displacement
correlations from our Bethe ansatz result in Eq. (\ref{Bethe_corr}).
Comparing Eq. (\ref{Bethe_corr}) and Eq. (\ref{d-u-rel}), we deduce
\begin{equation}
\label{uu-bethe}
[\langle\partial_x u_\alpha(x,y)(\partial_x
u_\alpha(0,0)-\partial_x u_\beta(0,0))\rangle] = 
\frac{\partial^2}{\partial x^2}
\frac{a^2}{2\pi}\sqrt{l_d \rho}\,\ln\left(\frac{x^2}{a^2}+
\frac{y^2}{\xi_\|^2} \right)^{1/2}.
\end{equation}
To complete the comparison with RG and VA, the elastic constants have
to be determined in the low density limit $\rho \ll 1/l_d$, where the
above prediction holds. The tilt modulus is given by $c_{44}=g/a$,
whereas the compression modulus $c_{11}$ is determined by the mutual
interaction potential $W(x)$ of the lines. Due to entropic and steric
repulsions, the original contact potential becomes long-ranged. The
interaction potential can be read off from the free energy density
$f(a=1/\rho)$ since in a grand-canonical description with chemical
potential $\mu$ one has
\begin{equation}
f(a)=\{-g(\mu/\mu_c-1)+W(a)\}/a.
\end{equation}
The elastic constant $c_{11}$ is now determined by the curvature of
the free energy density, evaluated at the mean line separation $a$, as
\begin{equation}
c_{11}=x^2f''(x)|_{x=a}=aW''(a)=\frac{\Delta}{Ta^2}.
\end{equation} 
To calculate the last expression, which is valid in the low density
limit $a \gg l_d$, we have used the result for the free energy in
Eq. (\ref{free-energy}), which yields the non-trivial part $W(a)/a$ of
the free energy density. From this result for the elastic constants we
obtain $J=(ag\Delta/T)^{1/2}/4\pi^2$, and hence the amplitude of the
$\ln$ term of the RG result in Eq. (\ref{RG_corr}) becomes 
\begin{equation}
\frac{a^2T}{8\pi^3 J}=\frac{a^2}{2\pi}\sqrt{l_d \rho},
\end{equation}
which is in complete agreement with the Bethe ansatz result of
Eq. (\ref{uu-bethe}). Moreover, there is also agreement between the
two results regarding the anisotropy parameter, since
$\xi_\|=(gT/\rho^3\Delta)^{1/2}=a(c_{44}/c_{11})^{1/2}$ in the dilute
limit. This means that the velocity of sound is related to the elastic
constants by the mapping between bosons and replicated lines according
to
\begin{equation}
\left(\frac{c_{11}}{c_{44}}\right)^{1/2} \longleftrightarrow v_s.
\end{equation}
It is interesting to note that exactly the same relation holds in the
case of a {\em pure} $1+1$-dimensional line lattice, which can be
mapped to world-lines of {\em free} fermions \cite{Pokrovsky79} with
$v_s=\pi\hbar\rho/m \leftrightarrow \pi T \rho/g$. 

Compared to the VA, our Bethe ansatz result for the difference
correlations $[\langle\partial_x u_\alpha(x,y)(\partial_x
u_\alpha(0,0)-\partial_x u_\beta(0,0))\rangle]$ is only in agreement
with the replica symmetric case. A universal amplitude of the
logarithm as predicted by the VA with replica symmetry breaking is not
reproduced by our approach.  However, the agreement of the Bethe
ansatz result with the RG predictions is not complete. If the RG is
assumed to give the correct result, then there should appear also a
slower decaying contribution proportional to $\ln(r)/r^2$ in the Bose
gas correlations in Eq. (\ref{bose_corr_1}) which remains finite in
the $n \to 0$ limit. This contribution corresponds to the last
off-diagonal term in Eq. (\ref{bose_corr_2}), for which this slower
deacy follows from the RG result in Eq. (\ref{RG_corr_2}).

There are two potential reasons for the absence of such kind of
contributions in our Bethe ansatz approach. The first one is not
related to our mapping, but can be traced back to the fact that there
are indeed logarithmic corrections to Haldane's harmonic fluid
approach for the 1D Bose gas. These corrections vanish in the strong
coupling limit $c_b\to\infty$ and have been calculated explicitly so
far only in a $1/c_b$ expansion by Korepin \cite{Korepin84} using the
quantum inverse scattering method and by Berkovich and Murthy
\cite{Berkovich89,Berkovich91} by means of conformal field
theory. However, these results predict, in agreement with each other
to first order in $1/c_b$, logarithmic corrections only to the
oscillating term of Haldane's result in Eq. (\ref{bose_corr_1}).
Therefore these results do not account for $\ln$ contributions to the
homogeneous part of the density-density correlation function. But we
are interested in the weak coupling limit $c_b\sim n^2 \to
0$, where the actual form of the corrections is not yet known. The
second potential, and we think more likely reason for the absence of
corrections is related to the fact that we apply our mapping to obtain
the excitation spectra of the $U(n)$ fermions from the Bose gas
spectra by assuming that only non cluster breaking replica symmetric
excitations are gap-less. This assumption is certainly justified for
integer $n$ due to the finite energy necessary to break up a
cluster. But if $n$ goes to zero, cluster breaking excitations might
become gap-less, providing a possible source of additional
contributions to the asymptotic density-density correlations. Note
that these cluster breaking excitations can be absolutely replica
symmetric since $n$ particles of mutual different colors can be
transfered from different clusters across the Fermi surface to form
together a color neutral excitation. Taking into account this type of
excitations is, of course, very interesting\footnote{The splitting of
clusters has been considered by Parisi for a single line ($N=1$) to
study the possibility of replica symmetry breaking \cite{Parisi90}.},
but not possible within the mapping to bosons, which itself cannot be
broken apart.

Finally, we come back to the oscillating term in the Bose gas
correlation function of Eq. (\ref{bose_corr_1}), and its implications
for the correlations in the line lattice. This term and higher
harmonics are quite important in view of experimental determinations
of the translational order of line lattices. Scattering experiments
measure the Fourier transform of the density-density correlation
function around a reciprocal lattice vector $2\pi m \rho$ with integer
$m$. Therefore, these experiments can determine the value of the line
lattice analogue $\eta$, of the exponent $\eta_b$ of the Bose gas
correlations. Using the mapping between lines and bosons, we have seen
that $\eta_b \sim 1/n$ [see Eq. (\ref{eta})] and therefore
\begin{equation}
\eta\to\infty,
\end{equation}
in the $n\to 0$ limit. This result is in strong contrast to the case
of a {\em pure} line lattice with $\eta=2$. The infinite exponent
means that the oscillating terms in the density correlation function
decay faster than any power law due to destruction of quasi-long-range
order by the random potential. Exactly the same result was obtained by
Cardy and Ostlund in their RG approach \cite{Cardy82}, but it is in
contradiction to all VA approaches.  Although a mean squared relative
displacement which grows slightly faster than a logarithmic increase
cannot be obtained directly from the homogeneous part of the Bose gas
density correlations, the fact that the Bose gas exponent $\eta_b$ is
mapped onto an infinite $\eta$ for the line lattice might be
interpreted as a signature of the presence of these
contributions. However, this argument has to be viewed with care since
$\eta\sim 1/n$ means that the oscillating terms of the density-density
correlation function have an essential singularity in the $n\to 0$
limit.

\section{Conclusion and discussion}

In this paper, we studied the probability distribution function of
the free energy of a lattice of lines, distorted due to thermal and
disorder fluctuations. Using the replica approach in combination with
an analytically continued nested Bethe ansatz, we 
calculate exactly the cumulants of the free energy in the dilute
limit. Our results confirm a previously proposed scaling form for the
replica free energy in terms of the the replica number $n$, as a
scaling field \cite{Emig00}. These exact results were obtained by
establishing a novel analogy between the random line lattice and the
weakly interacting 1D Bose gas with delta-function interactions.
Based on this analogy, we related the results for the density-density
correlation function of the Bose gas, known from Haldane's effective
theory of 1D quantum liquids, and from conformal field theory, to
correlation functions of the line lattice.  Our result for the
difference between diagonal and off-diagonal replica-correlations is
in agreement with RG predictions, and a replica symmetric variational
ansatz for an elastic model of the line lattice, if one takes into
account the correct disorder dependence of elastic constants
obtained by the Bethe ansatz approach for the free energy. The
Bethe ansatz results do not agree with the variational ansatz with
replica symmetry breaking.  The comparison with the RG results of
Cardy and Ostlund \cite{Cardy82} and Goldschmidt and Houghton
\cite{Goldschmidt82}, which predict a faster than logarithmic growth
of relative line displacements, shows that our mapping to a Bose gas
cannot capture the corresponding contributions.  Potential sources for
these additional logarithmic terms within the Bethe ansatz provide
avenues for future explorations.

Finally, it might be interesting to note that all results obtained in
this paper are valid if the length scale $l_d$, setting the crossover
to random behavior for a single line, is much larger than the
correlation length $\xi_0$ of the random potential. When this
condition is fulfilled, $\xi_0$ does not show up in the final results,
and can be safely set to zero from the beginning as we did throughout
the paper, see Eq. (\ref{randpot_corr}). However, if temperature is
decreased, the length scale $l_d=T^3/g\Delta$ also decreases, and gets
equal to $\xi_0$ at the crossover temperature
$T^*=(\xi_0g\Delta)^{1/3}$, which is of the order of the smallest
random energy barrier \cite{Nattermann88}. Therefore, our results
apply directly to the high temperature regime $T \gg T^*$. They can
also be extended to the low temperature limit $T \ll T^*$, using a
method developed by Korshunov and Dotsenko \cite{Korshunov98} to
calculate the replica free energy of a single line in a random
potential with finite $\xi_0$. By splitting the $n$-cluster for a
single line into $n/k$ separate blocks with $k\sim T/T^*$, they were
able to show that the ground state energy of their model with finite
$\xi_0$ can be obtained from the model with $\delta$-function
interactions by the simple substitutions $g\to kg$, $\Delta\to
k^2\Delta$ and $n\to n/k$. Since the internal structure of individual
clusters remains unchanged by considering a finite density of clusters
(see the ans\"atze in Eqs. (\ref{lam_ansatz_1})-(\ref{lam_ansatz_3})),
our results can be translated easily to the low-temperature regime by
the above substitutions.

\section*{Acknowledgements}

\begin{sloppypar}
We would like to thank E.~Frey, A.~Hanke, V.~E.~Korepin,
T.~Nattermann, V.~L.~Prokovsky and W.~Zwerger for discussions and
conversations.  This work was supported by the Deutsche
Forschungsgemeinschaft under grant No.~EM70/1-3 (T.E.), and by the
National Science Foundation grants No.~DMR-98-05833 and
No.~PHY99-07949 (M.K.).
\end{sloppypar}

\end{document}